%% file: ConsensualVeracity.tex
\documentclass[a4,twocolumn,10pt]{article}
\usepackage{usenix2019_v3} 
\usepackage{url}
\usepackage{tikz}
\usepackage{amsmath,scalerel}
\usepackage{amsthm}
\usepackage{filecontents}
\usepackage{graphicx}
\usepackage{amssymb}
\usepackage{epstopdf}
\usepackage{caption} 
\usepackage{booktabs}

\usepackage[scaled=0.88]{helvet}  	
\usepackage[T1]{fontenc}
\usepackage{float}
\usepackage{amssymb}
\usepackage{cuted}			
\usepackage{listings}
\usepackage{graphicx}
\usepackage{subfigure}
\usepackage{adjustbox}
\usepackage{xspace}
\usepackage{cite}
\usepackage{tabularx}
\usepackage{enumitem}
\usepackage{threeparttable}
\usepackage{enumitem}
\usepackage{mathrsfs}
\usepackage[ruled]{algorithm}
\usepackage[noend]{algpseudocode}
\usepackage{mathtools}
\usepackage[n,operators,advantage,sets,adversary,landau,probability,notions,logic,ff,mm,primitives,events,complexity,asymptotics,keys]{cryptocode}
\usepackage{comment}
\excludecomment{confversion}\includecomment{fullversion}

\begin{document}
\input{macros.tex}

\title{\Large \bf Distributed Governance  \\ a Principal-Agent approach to data governance \\ \sl{Part 1 background \& core definitions}} 

\date{August 2023}

\author{
    {Paul Knowles} \\
    \and 
    {Philippe Page \thanks{Corresponding author, \href{mailto:philippe.page@humancolossus.org}{philippe.page@humancolossus.org}}}\\  \\ Human Colossus Foundation \\Geneva, Switzerland
    \and 
    {Robert Mitwicki} \\ 
} 

\maketitle
\input{CV0-abstract.tex}
\setcounter{tocdepth}{2} \tableofcontents
\input{CV1-introduction.tex}

\input{CV2-DistrGover.tex}
\input{CV3-ConsensualVeracity.tex}
\input{CV4-Conclusions.tex}

\nocite{*} 
\bibliographystyle{ieeetr}
\bibliography{CVeracity-References20230718}
\end{document}

%% file: macros.tex
%
%
%
\newcounter{subdefinition}
\renewcommand{\thesubdefinition}{\thedefinition.\arabic{subdefinition}}
\newenvironment{subdefinition}{
        \refstepcounter{subdefinition}
        \par\noindent
        \textbf{\upshape Definition \thesubdefinition}%
}{}

\setlength\marginparwidth{16mm}
\setlength\marginparsep{1mm}
\setlength\marginparpush{2mm}
\newcommand{\commentmarker}[1]{\colorbox{red}{\textcolor{white}{#1}}}
\newcounter{CommentCounter}
\newcommand{\commentUp}[3][]{%
  \stepcounter{CommentCounter}%
  {\scriptsize\commentmarker{\theCommentCounter}}%
  \marginpar{%
    \vspace{-#2}
    \tiny\raggedright
    \commentmarker{\theCommentCounter}
    \ifstrempty{#1}{}{\textcolor{red}{#1: }}%
    {\scriptsize #3}%
  }%
}
\newcommand{\internalcomment}[2][]{\commentUp[#1]{0pt}{#2}}

\newcommand{\name}{\textsc{DDE}\xspace}
\newcommand{\nameF}{\textsc{Dynamic Data Economy}\xspace}
\newcommand{\DDE}{\name}
\newcommand{\DDEFull}{\nameF}
\newcommand{\mycomment}[1]{}

\theoremstyle{definition}
\newtheorem{definition}{Definition}
\newtheorem{protocol}{Protocol}
\newtheorem{lemma}{Lemma}
\newtheorem{theorem}{Theorem}
\newtheorem{axiom}{Axiom}					
\newtheorem{subDefinition}{Definition}[definition]	

\newcommand{\parab}[1]{\vspace{1mm}\noindent\textbf{#1}}          
\newcommand{\parait}[1]{\noindent\textit{#1}}                     
\newcommand{\paraNormal}[1]{\vspace{1mm}\noindent\normalfont{#1}} 
\newcommand{\textvar}[1]{\textrm{\textsf{#1}}}

%% file: CV0-abstract.tex

\begin{strip}   
\vspace{-1cm}
\begin{abstract}
To address the need for regulating digital technologies without hampering innovation or pre-digital transformation regulatory frameworks, we provide a model to evolve {\sl Data governance} toward {\sl Information }governance and precise the relation between these two terms. This model bridges digital and non-digital information exchange.
\par\quad
By considering the question of governed data usage through the angle of the {\sl Principal-Agent problem}, we build a distributed governance model based on {\sl Autonomous Principals} defined as entities capable of choice, therefore capable of exercising a transactional sovereignty. Extending the legal concept of the privacy sphere to a functional equivalent in the digital space leads to the construction of a {\sl digital self} to which rights and accountability can be attached.  Ecosystems, defined as communities of autonomous principals bound by a legitimate authority, provide the basis of interacting structures of increasing complexity endowed with a self-replicating property that mirrors physical world governance systems.
\par\quad
The model proposes a governance concept for multi-stakeholder information systems operating across jurisdictions. Using recent software engineering advances in decentralised authentication and semantics, we provide a framework, {\sl Dynamic Data Economy} to deploy a distributed governance model embedding checks and balance between human and technological governance. Domain specific governance models are left for further publications. Similarly, the technical questions related to the connection between a digital-self and its physical world controller (e.g biometric binding) will be treated in upcoming publications.  
\par\quad
\end{abstract}
\end{strip} 

%% file: CV1-introduction.tex
%
\section{Introduction}
\label{sec:introduction}
A hasty digital transformation undermining society's fundamental need for protection to secure their values justifies the need for new governance models. Despite a high rate of technological innovations, digital transformation faces adoption challenges virtually everywhere. Fears or mistrust either delay innovations or confine them within the platform's walled gardens, limiting their potential reach to all citizens. In other words, the security of information systems is the root cause of most problems related to data usage.
\par
Cybersecurity is a mature industry protecting data from theft and hacking. However, more is needed. Data usage also requires protection, which is a much more complex problem. Securing data amounts to securing a stream of bytes processed or recorded by a system. However, securing data usage requires, in addition, the knowledge of the context. The context gives meaning to data, and this is where the true value of data resides.
\par
This work proposes a model to evolve the current data governance mechanisms to consider the human context. Defining {\sl information} as the combination of data and meaning, we bring the problem in the domain of {\sl information governance} that is not limited to the digital space. 
\par
As a result, {\sl security} needs to be considered in a broader sense and encompass existing legal and ethical frameworks that govern our communities. Furthermore, at the societal level, the security of the individuals and the community goes beyond physical security and relies on these trust frameworks that we will label governance in this article. As a result, to ensure that security is available to all systems, we argue for a digital technology that shifts core security from the applications to the network protocols. 
\par
Information governance also impacts the economics of data. Currently, the data itself is valued, not its usage. The centralisation of data is akin to the centralisation of wealth. By considering the context, information governance introduces a form of digital accountability through the concept of the digital self. As such, the model can also be used to design value exchange mechanisms attached to data usage.
\par  
The introduction to the distributed governance model starts with a call to urgently address digital transformation in the humanitarian sector, where data security has to protect vulnerable populations when Human governance cannot. The remainder of the introduction provides further information on the rationales underlying the need for distributed governance models. Section \ref{subsec:socialImpact} focuses on how the growing mistrust in information systems relates to the platforms model's technological limitations. Despite the enormous efforts made by the public and private sectors to solve the privacy, surveillance and dependency problems related to information centralisation, the trend towards further mega-monolithic platforms cannot be curbed (at least economically). Section \ref{subsec:alternative} relates mistrust in information systems to the {\sl agency dilemma} and the Principal-Agent approach. The introduction ends with section \ref{subsec:security} where we highlight technological complexities that make digital technologies hard to govern. For example, how do we assign liability in a digital space?

\subsection{The urgency to act\\{\sl Digital transformation for Humanitarian action}}
\label{subsec:humanitarian}
\begin{center}
\begin{figure}[h]
    \includegraphics[width=80mm ]
    {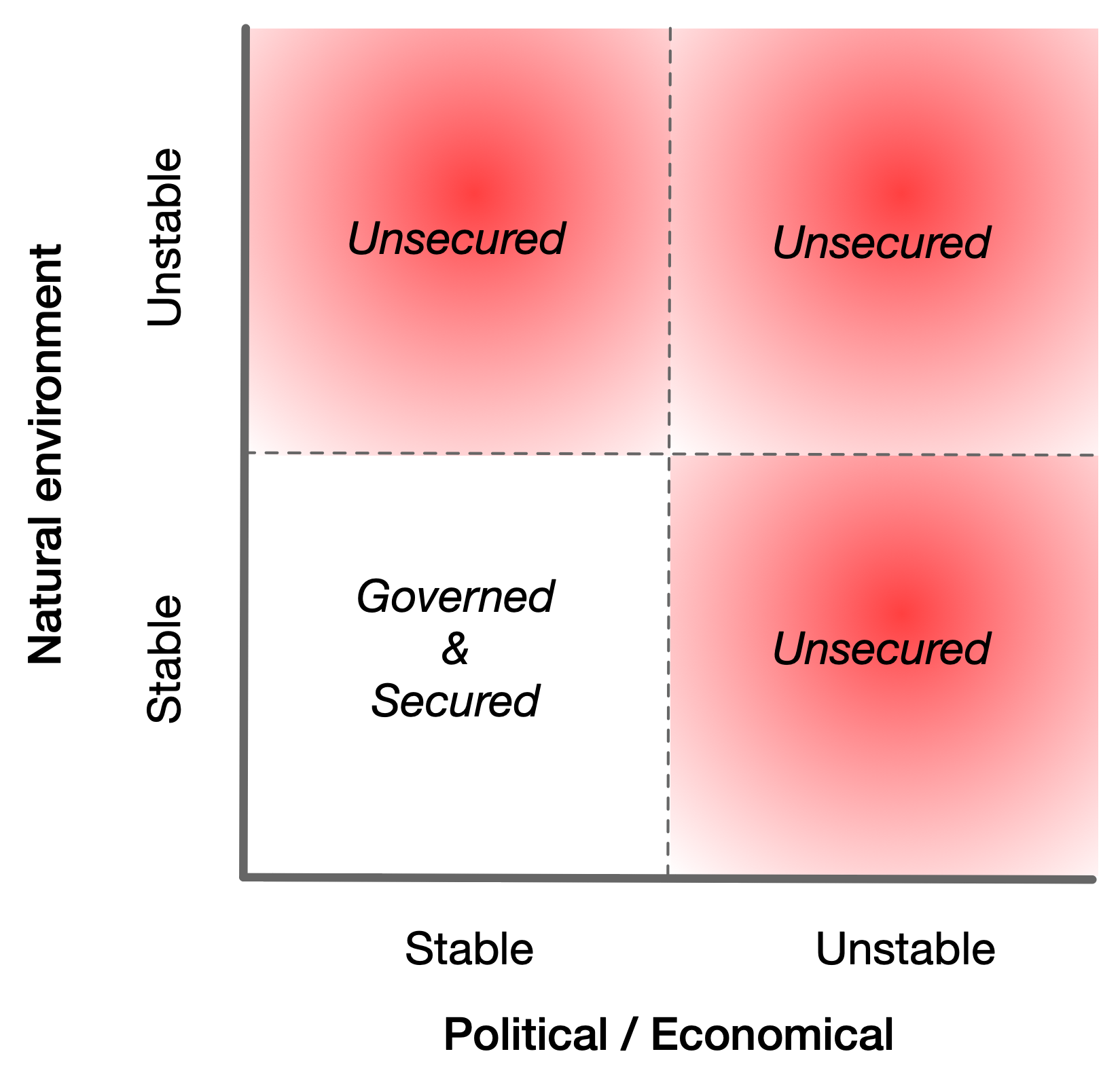}
    \caption{Security designed for unstable environments}
    \label{fig:caseForHumanitarian}
\end{figure}
\end{center}

Digital technologies have the potential to provide a safe harbour for vulnerable population in a crisis. They are even recognised as a necessary element in the new balance of risk facing humanitarian action \cite{daccord2016}. But for technologies to help people in need, its usage has to be intrinsically secured for the dependent person and avoid further increasing their dependency on external resource. Most of the current efforts to secure the digital space are applicable to stable conditions while humanitarian action always takes place in unstable conditions. Figure \ref{fig:caseForHumanitarian} calls for action to develops mechanisms that secure information usage in humanitarian action.
\par
Instabilities can be due to natural causes (earthquakes, drought, floods), political tensions (e.g. war, regime change) or economic mayhem (e.g. poverty, miss management). In all cases, the troubles reduce the security provided to the individual by their community and institutions. As a result, disrupted information flows can not be secured and further increase the vulnerability of populations or reduce the efficiency of external help. Therefore information security must be integrated into the protocols when legal protection is unavailable.
\par
This call for protocols that secure the data itself on top of the existing security provided by the platforms or services. The intrinsic benefits of digital information, instantaneous communication of any media (text, sound, video), and global reach are paramount to managing a crisis. Therefore, core network protocols themselves must bear a greater role in securing the network independently of the applications connecting to the network. Technologist call this trend the move from the current {\sl fat applications, thin protocols} to {\sl thin applications, fat protocols}. 
\par
Before these benefits can materialise, the concerned individuals must trust the information system. Therefore the distributed governance model must also take into account the protection of individuals in crisis when traditional protection mechanisms do not protect them anymore. Thus the model is design around the existence of decentralised software technologies that can provide decentralised control to the human when the context requires it.
\par
The work of ICRC on data protection \cite{KUN:2017,KUN:2020} or the Oxford Internet Institute on the usage of decentralised technologies  \cite{Cheesman:2020, Cheesman:2021, Martin:2022} shows that the Humanitarian sector is well aware of the inherent risks in digital information systems. Decentralised ledger technologies (blockchain) and biometrics are also introducing the risks of dependency mentioned in the opening paragraph of this section. Digital security in the Humanitarian sector requires governing technology at a time and place where Human governance has failed or is unreliable. Is it doable?
\par
This work on distributed governance points to an alternative usage of todays network where data is treated as first class citizen and data usage governed by mechanisms derived from existing governances embracing new technologies, not directed by new technologies.
\par
In conclusion, addressing data governance questions for specific humanitarian action where human governance is unstable will bring the essential elements needed to provide security. Integrating them into stable governance will be easier and might be an accelerator of digital transformation for society as whole 

\subsection{The limits of digital platforms\\ {\sl Intrusive digital centralisation}}
\label{subsec:socialImpact}
At the scale of humanity, the spread of digital transformation took place at a fantastic speed in little more than a generation. Over half of the world's population connects to the Internet. Digital transformation is now a crucial enabler of the world economy and necessary to meet the UN sustainable development goals (SDG)\cite{UN:2017}. In 2019, the UN secretary general published {\sl "The Age of Interdependence"} \cite{UN:2019b}, a report calling for cooperation at all levels of society in the digital space. It is a potent reminder of the unmatched capacity of humans, as a species, to cooperate on a large scale. Therefore, recognising the fundamental role of cooperation in human nature and societal developments provides a new angle on the protection mechanisms (i.e. security) to be implemented. 
\par
Today, digital networks continuously connect over half of the world's population, organisations, and governments. If we add the growing number of online monitoring devices, the networks become a "digital space", a fabric similar to our physical space where we extend our activities. Through digital transformation, we live in and depend on a hybrid physical-digital space. In particular, the societal questions raised by digital transformation show its impacts on the equilibrium between the protection of the self (i.e. individual) and the protection of the community (i.e. society). Therefore, the protection mechanisms (i.e. governance) must be coherently extended to this hybrid space.
\par
Currently, platforms, as the network's nodes providing services, are the primary interaction system for most users. Platforms have expanded to become quasi-ubiquitous and are at the centre of a hyper-connected society. We point here to the limitations of platforms that motivate introducing a distributed governance model. The section \ref{subsec:security} will deal with the problem's root cause and indicates implementation paths for distributed governance models.
\par
Efficient platforms result from the compelling capacity of Service-oriented Architecture (SOA), where user requests are processed, and results returned in an appealing form. The success of platforms to bring innovations to life are too numerous to be listed. However, these successes come with a price, the increasing demand for data from the platforms to perform the services requested by users. With data perceived as the fuel of services, data handling becomes a process second to the service controlled by each platform. As a result, data processing left to the platforms leads to obscure operations performed behind closed walls. 
\par
As technologies to capture data have improved, tsunamis of data are constantly flooding platforms, creating severe problems for both platforms and users. These problems highlight two limitations of the current platform models.
\subsubsection*{data processing and fundamental rights}
First, we consider the impact of data processing within platforms on fundamental values. The economic incentives for platforms to provide more value-added services translate into a growing appetite to collect more data. Therefore, this generates a vicious circle. By increasing platforms' liabilities to secure data, costs increase; therefore, the pressure to process data outside the initial intended purposes also increases. Ultimately this further fuels the demand for data. End users' consequences can be expressed as a Cornelian choice between fundamental values and immediate benefits. The user's incentives to trade the long-term risks of data misuse for the short-term gains of the services prevent an educated or rational decision.
This problem of choice, data vs service, is present at scales beyond decisions made at individual levels. Healthcare provides an example of a domain where the potential misuse of data significantly slows down the development of new digital solutions across the whole ecosystem. If advanced goals to provide personalised medicine are becoming technological realities, they also bring the systemic risk of data misuse. The European Commission's concept of a European Health Data Space (EHDS) launched in April 2022, as well as major policy initiatives like the Data Governance Act (2020) or the Artificial Intelligence Act (2021), demonstrates the urgency to find solutions and the importance of the related challenges. Similar efforts are taking place worldwide, each reflecting their approaches to balance their fundamental values with the benefits arising from digital innovations. 
\subsubsection*{accuracy of data processing}
A second limitation of the data processing behind the platform's walls is accuracy. The increased power of digital analytical tools comes with the requirement to ensure precise and robust control over the input data sets. This limitation is significant for the sub-field of artificial intelligence called Machine Learning (ML) where data sets serve as training input for algorithmic processing. By design, ML means learning through patterns without understanding the underlying mechanisms. Therefore improving the output's accuracy translates into higher demand for larger input data sets and more precision in the lineage (i.e. the deterministic origin of data items). A related but distinct limitation is the cost of correlation when contextual information materially impacts the outcome of an analytical tool. For example, timely access to additional information poses a technical, regulatory and sometimes reputational challenge for the analytical tools platforms. In response to these limitations, we observe the resurgence of Data-oriented Architecture (DOA) to go beyond service-oriented architectures' barriers.
\par
In conclusion, the current trend of growing {\sl data rich} platforms leads to the centralisation of data processes that are bound to become more intrusive and impact fundamental rights of communities to meet output or results expectations. To overcome the limitations of the current platforms, more efficient mechanisms to control the data outside platforms are needed. In other words, data is not to be understood as the fuel of a digital service but as an object defined and protected independently of the platform processing it.

This conclusion is a requirement for the design of a distributed governance model.
\subsection{On governance and liability\\{\sl Securing information Systems}}
\label{subsec:security}
\subsubsection*{an anthropological argument}
\begin{center}
\begin{figure*}[h]
 \includegraphics[width=\linewidth,trim=3mm 3mm 2mm 0mm]
{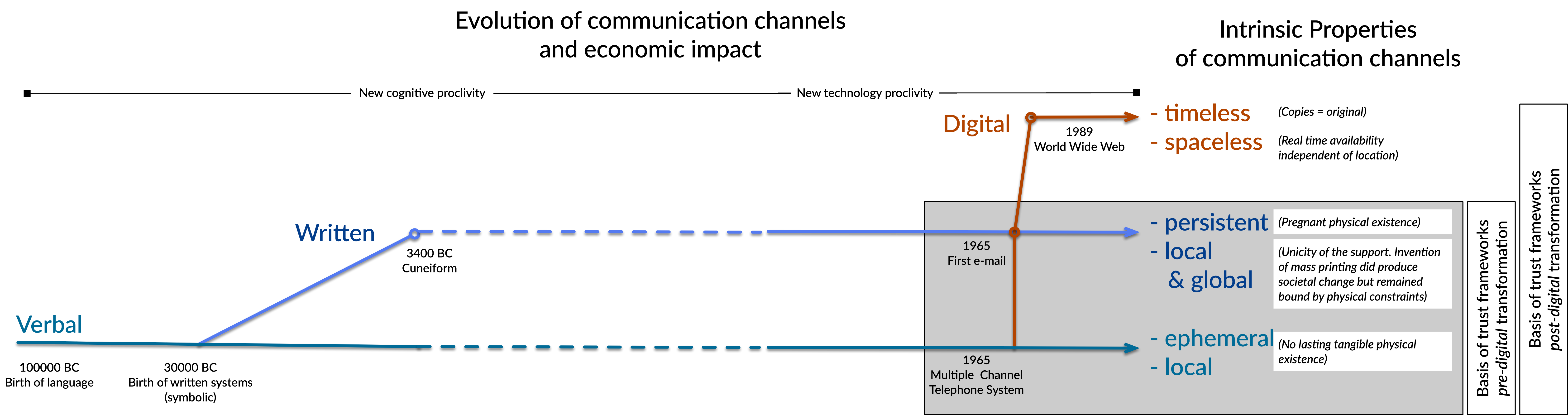}
 \caption{HCF perspective on Cognitive vs Technology proclivity}
\label{fig:behaviorChange}
\end{figure*}
\end{center}

We base our governance work on the observation that digital transformation has radically changed how society operates at all levels. The impact goes beyond technology and should be considered from an anthropological perspective. Digital information processing is a quantum leap from traditional oral and non-digital written information processing. The intrinsic properties of digital information are significantly different from the intrinsic properties of written and verbal communications that were, until recently, our sole basis for building trust frameworks (governances) to secure communities and their members. By considering the evolution of communication alongside the significant organisational advances of humans  \cite{Tattersall:2016}, it becomes clear ({\sl see figure \ref{fig:behaviorChange} on page \pageref{fig:behaviorChange} }) that the new trust frameworks must extend the existing trust frameworks to include communication channels that are not bound by the physical constraints of time and space as written and verbal communication are. 
Two critical intrinsic properties of digital information, not found in pre-digital transformation, are related to how information propagates through time and space. First, digital information is {\sl timeless}. Any byte or file (i.e. a group of bytes) in a digital system can be copied multiple times, and nothing will distinguish the original from a copy. Time stamping mechanisms or contextual information have to be added on top. As a result, by itself, data does not age. Second, digital information is {\sl spaceless}. It can be located anywhere and in multiple places at once. These two fundamental intrinsic properties of digital information have profound importance when designing a governance framework. Taken together, they lead to the loss of the unicity of digital information. This unicity is essential to determine what governance applies to and who is liable.
\par 
The link to information security is through governance. More precisely, human society started from individuals gathered around increasingly complex communities of interest to protect their values (commercial, ethical, religious). We continuously develop structures defending the community against external and internal threats and ensure the group's efficient and coherent development. From the initial hunter-tribes to today's complex mesh of sovereign nations interlaced with global organisations (public or private), the same security pattern is at work. A community can be described as a consensus of members around a set of codified rules (e.g. constitution) or generally accepted values (e.g. ethics). Security evolved from the individuals to structures centralising protection (e.g. an army, a legal framework).
\par 
Securing societies is the evolutionary driver behind the increasingly complex mesh of rules governing the actions of all economic agents, individuals, public or private alike. Physical and legal structures protect our values as individuals, organisations or nations. Security starts with autonomous individuals growing communities based on interests or necessity.  From the family cell, the corporation, and a country up to the United Nations, each community centres around a common set of rules (legal or ethical). 
\par 
Our approach to information system security based on community-defined governance allows a more detailed consideration of individual privacy.  Privacy as a fundamental right protected by the community helps to give individual privacy a societal dimension ,as advocated by Mokrosinska \cite{Mokrosinska:2018}.
 \subsubsection*{Technology security vs Society security }
\begin{center}
\begin{figure}[h]
    \includegraphics[width=80mm ]
    {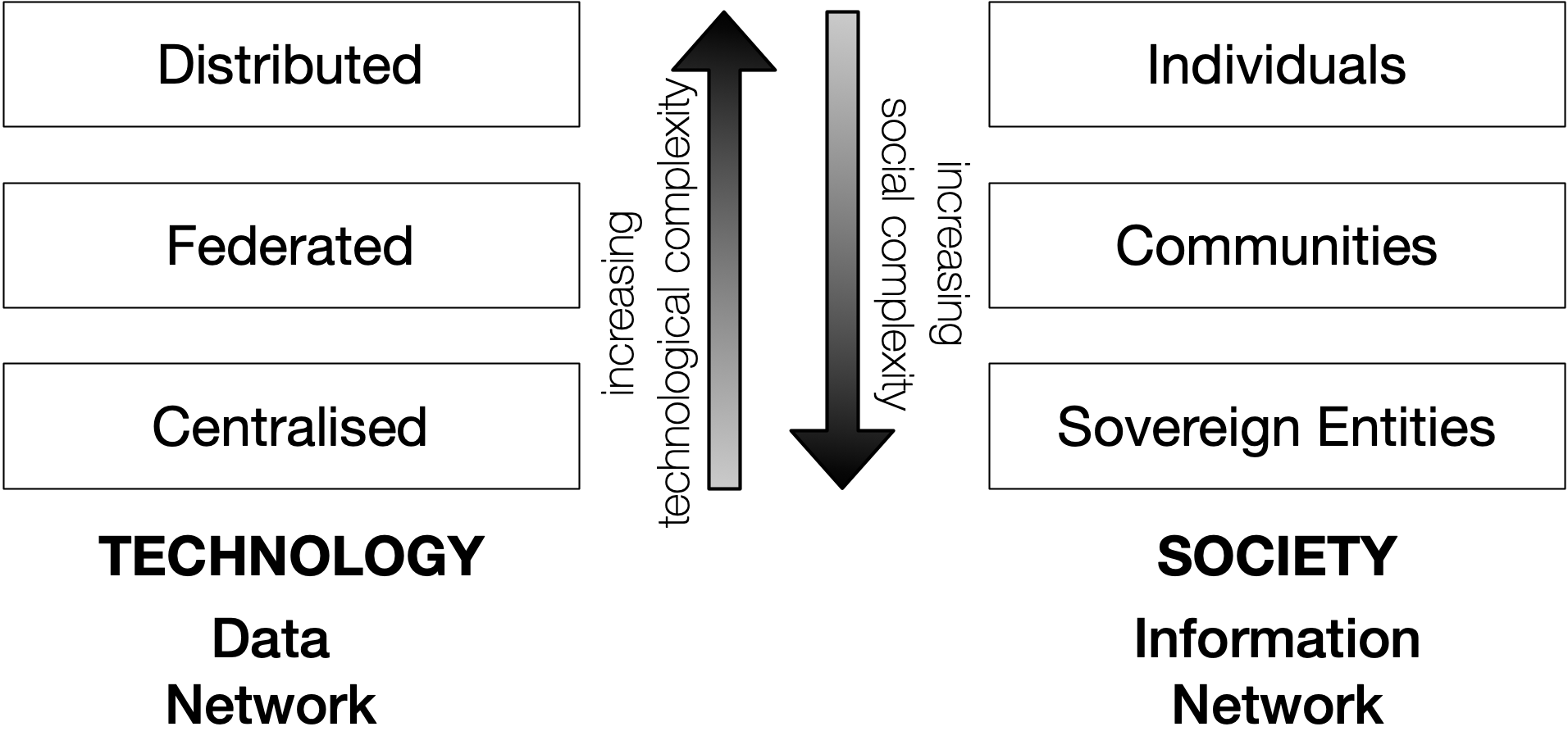}
    \caption{The {\sl Society} vs. {\sl Technology} security governance evolution}
    \label{fig:technologySociety}
\end{figure}
\end{center}

 Digital security took the opposite path, as summarised in figure \ref{fig:technologySociety} on page \pageref{fig:technologySociety}. Security of the communication network drove the development of the Internet. Information technologies started with centralised systems, and the Internet evolved into distributed ecosystems\cite{Baran:1964}.The results of the transition went beyond the expectations of the pioneers. Today, with the advent of the World Wide Web open protocols in 1989 \cite{Berner1990}, the Internet has become the backbone of a planetary mesh connecting billions of individual users, organisations and a growing number of things. 
\begin{center}
\begin{figure}[h]
    \includegraphics[width=80mm,
    ]
    {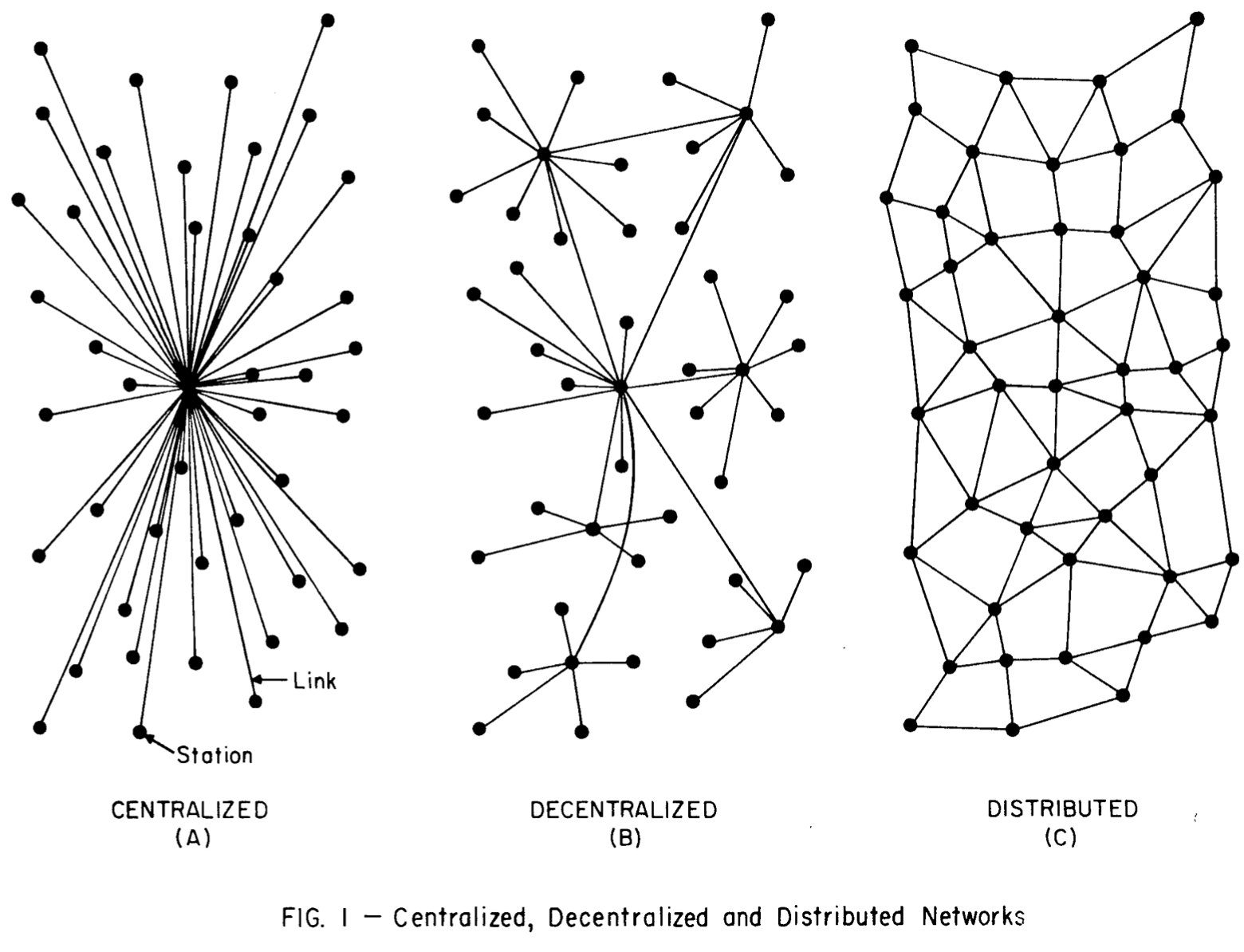}
    \caption{Network security driven by decentralisation. Source, Rand corporation  \cite{Baran:1964}}
    \label{fig:internetModel}
\end{figure}
\end{center}
Thus contrary to information systems, security starts from distributed entities that agree on centralising governance for their protection. Thus securing societies goes from distributed to centralised while securing information systems historically went from centralised to distributed due to technological complexities, as represented in figure \ref{fig:technologySociety}
The technology and society approach to security reconciles if the existing governance models extend their base to include the intrinsic properties of digital communication. This extension is within reach of software technologies with the recent advances in decentralised software technologies for authentication and semantics. 
\par The following two aspects will be described in more details in the last section of this report. First, decentralised semantics aims to provide pairwise digital object integrity. As a result, two network agents can agree on the meaning of digital objects without needing a third-party intermediary. Therefore, peer-to-peer agreement on meaning is necessary to ensure the exchange of harmonised data when multiple stakeholders operate in multi-jurisdictions ecosystems.
\par
Second decentralised authentication is required. On open networks, it ensures the authentification of network actors independently of a technological third party. Often underestimated, decentralised authentication is also necessary to independently authenticates events. For example, time stamping events in a deterministic context ensure causality without relying on external parties.

This paper argues that recent advances in the design of two core software architectures of digital networks, authentication and semantics, provide an alternative to the current technology-driven top-down information security design. With information systems becoming increasingly intrusive (e.g. social media), the top-down digital technology security distorts societies' governance model by centralising decision and enforcement power within technologically driven actors. In addition, as digital technology is also a driver of economic growth, the development of trust frameworks clashes with the traditional bottom-up consensual development of community governances that must secure the fundamental liberties of the individuals, organisations and institutions protected by the existing trust frameworks.

The combination of these technologies builds the digital core elements of unicity (i.e. accountability) and context(i.e. severity) necessary in building governance. This four-part series describes an alternative information security model in the digital networks aligned with the existing structures that societies have established to develop trust. The first part details a concept of distributed governance that allows human trust frameworks to extend into the digital space. This extension is achieved by defining autonomous principals in section \ref{subsec:alternative}. We distinguish agents with Free Will from algorithmic agents that mechanically respond to the environment. Therefore, the model includes accountability and consent at its core. The advances in decentralised semantics and authentication only serve as technological tools to ensure the existence of deterministic digital objects and the authenticity of events for usage in existing rules and regulations.

Section \ref{sec:distributedGovernance} defines this hybrid governance model dealing with humans and machines. section \ref{sec:consensualVeracity}  details how the recent advances in decentralised semantics and authentication can be applied to enable a digital information network that can be tied to existing rules and governances across jurisdictions. In this report, we provide the introduction to the model and leave explicit construction for subsequent publications.
The remainder of the introduction provides background information on the topics later presented.

\subsection{The Principal-Agent approach\\ {\sl Governance of information systems}}
\label{subsec:alternative}
Ultimately, the distributed governance model presented here aims to define a trust architecture where the core software components federate cybersecurity and privacy protection efforts. We first observe that the main problem with intrusive information systems is the unintended usage of the information, not the system itself and its intended purpose. This observation is an instance of the {\sl Principal - Agent problem} \cite{Eisenhardt:1989} for digital information systems. Some research applied the principal-agent theory in the digital sector, but only some approached it as a methodology for digital transformation. The novelty of our approach is twofold. First, we phrase the problem from the perspective of the Agent's liability (governance) in general and not only from the privacy perspective. Second, we describe a framework to provide functional requirements for information systems.
\par On one side, the Principal is the data user or subject defined in data protection regulations like GDPR. Nevertheless, conversely, the Agent is the information system (or a component). Therefore, when dealing with the risk of misuse of information, securing the data should start with the agency question. How can the user trust that the ICTs perform their mandated task and not something else?
\par Initially defined in political or economic science, the Principal-Agent problem addresses how to ensure a mandated Agent performs in the Principal's best interests without being influenced by its benefits.
With digital transformation increasingly driving changes in our societies, individuals, organisations and governments all face a form of agency question when interacting with an information system.
For individuals connecting to a digital service or a platform, the agency question deals with using their data outside their primary purpose. The individual will value the benefits of the purpose, while the platform has strong economic incentives to value its business models. When re-using data for a different purpose, the platform creates an economic asymmetry in the value of the data exchange.
For businesses and organisations relying on platforms to carry out their core business, the economic asymmetry will take the form of dependencies, vendor locking or total dependence on a third party for their data.
\par 
For the governments, in addition to the increased reliance on intermediaries to carry out core administrative tasks, the economic asymmetry leads to a potential loss of sovereignty over their data through an inadequate choice of technological solutions. The Principal-Agent perspective also provides insights into global surveillance. Governing institutions, as agents of the legitimate authority of the ecosystem they represent, face an adoption challenge of ICT when they must convincingly demonstrate that new data usage respects the rights of their constituents. The debates around the Covid-19 pass are only one recent example.
\par
Bringing the agency perspective in an information system forces the model to unambiguously identify Principals on a network where humans and digital devices constantly mix. Devices are algorithmically controlled machines; thus, reducing their identification to authentication is possible. On the other hand, humans have free will, and their complex relationship with their environment defines their identity. 
\par Currently, information systems reduce the identification of humans to authentification through the concept of accounts \& logins. Without considering the context of the authentication, the wording {\sl Digital Identity} leads to confusion and is often a misnomer. Identity is more than authentication.
\par In order to define a governance framework for an information system in interactions with the existing regulatory and trust frameworks of an ecosystem (e.g. a community), section \ref{sec:distributedGovernance}  introduces the concept of "Distributed Governance", where humans are singled out as agents with free will, are the Principals. We define the term autonomic agents to represent an entity with "free will", i.e. a capacity to decide. The term covers both physical and digital entities. Our definition of the autonomic Agent differs from the traditional engineering definition to stress that freedom of choice is:
\begin{enumerate}
\item not algorithmic;
\item bound to the Agent's accountability.
\end{enumerate}
In addition, discussing information systems from an agency perspective will allow a natural extension of accountability to structures built by autonomic agents, the {\sl Ecosystems}. In an analogy with the physical world, humans are the fundamental particles, the nodes of an information network. Then, the higher-level constructs (businesses, states, professional organisations) correspond to atoms and molecules. In this analogy, information exchanged or recorded is akin to the binding forces. 
\par In summary, the Principal-Agent perspective provides a holistic method to secure the information both in the physical and digital space. Cybersecurity and governance become two aspects of a technological problem under the higher governance of societies' trust frameworks.

%% file: CV2-DistrGover.tex
%
\section{The Distributed Governance Model} 
\label{sec:distributedGovernance}
\subsubsection*{\sl Generalities} 
ThThe Distributed Governance Model (DGM) presented here is an abstract model designed to develop explicit governance frameworks rooted in existing Human governance and supporting the intrinsic properties of digital information systems as highlighted in section \ref{subsec:security}. The primary added value of the DGM approach is the design of information systems operating across multiple jurisdictions that can scale to many stakeholders.
\par
Today, data governance refers to the rules applied to the data exchange between connected stakeholders and the organisation that implements and enforces them. Data governance frameworks comprise contractual agreements and adherence to specific standards and regulations. However, data governance frameworks expressed in a techno-corporate language are unsuited for a large mesh of open networks with many stakeholders interacting dynamically. As a result, public institutions have developed separate regulatory mechanisms to ensure data governance framework also comply to public regulations. This lead to the necessary (but insufficient) increase of data protection regulations for example. 
\par
The distributed governance model differs from current data governance as it better captures the existing jurisdictional context of data exchange. DGM introduces a functional definition of {\sl liability} in the digital space. Therefore DGM allows the introduction of existing regulations in the digital space in addition to the data-specific regulations.
\par
By design, the distributed governance model is data-centric and is well suited for so-called Data-Oriented Architectures (DOA) where digital exchanges occur across jurisdictions or platforms-based information systems. Therefore it can be applied in different real-world environments involving multiple stakeholders on a large scale, see for example\cite{NUS:2022, EHDS_20220503}. The model also has application in AI to provide better control over the quality and usage of machine learning data sets \cite{cabrera2023realworld}. Regulatory, ethical, scientific and community (public) considerations must be considered coherently to drive adequate data governance.
\par
As a result, the DGM can also be used for defining policy engine in a Zero Trust architecture \cite{NIST_2020zt,NIST2020}. Using the Dynamic Data Economy introduced in the last part of this work in section \ref{sec:consensualVeracity}, the model also provide the governance architecture for zero-trust {\sl environments} where independent platforms have to interacts within multiple jurisdictions.
\subsubsection*{\sl Core elements}
The first section \ref{subSec:autonomousPrincipals} define {\sl\bf Autonomous Principals} as entities endowed with a capacity of choice (i.e.free will). Autonomous principals are thus accountable for their actions. They can be used to describe humans, organisations or sovereign nations and are at the core of the definition of ecosystems and provide a representation of the carrier of trust and consent between ecosystems.
\par
Then section \ref{subSec:ecosystems} defines the {\sl\bf Ecosystem} as a functional representation of a community; a group of autonomous Principals gathered around a legitimate authority. We also differentiate between intrinsic and extrinsic properties of data leading to the concept of {\sl Information Network} where governances are anchored versus the {\sl Data Network} where intrinsic properties of data are secured.
\par
Sections  \ref{subSec:extrinsic} and \ref{subSec:confidentialitySpheres} defines the core extrinsic properties upon which privacy, consent and accountability can be build. {\sl\bf Confidentiality spheres}, as a functional extension of the legal concept of privacy sphere. Confidentiality spheres are attached to all autonomous principals. They represent the ultimate control of the principal over his information within and across ecosystems. 
\par
Within an ecosystem, a governance administration defines the rules of communication between autonomous principals. Across ecosystems, the autonomous principals provide the information portability mechanism reflecting the trust framework developed by the ecosystem themselves.
\par
We end with three examples to illustrate how the model can be applied to design solution in multi-stakholder environment. Explicit constructions are left to the subsequent publications in this series where specific examples will be detailed.
\par

{\sl (a notation summary is found in table \ref{table:notationAbstractGov} on page \pageref{table:notationAbstractGov})}
%

\subsection{Autonomous Principals\\{\sl Towards a definition of digital-self}}
\label{subSec:autonomousPrincipals}
Cooperation starts with a peer-to-peer relationships.Therefore the distributed governance model begins with actors endowed with the capacity of choice (i.e {\sl free will}) to cooperate or not, to transact or not. Governance is linked to the autonomous Principal's freedom of choice through accountability for the choices made. The definition below does not refer specifically to digital. As a result, it refers  both in the physical world and digital space. This is a necessity for any model that aims to protect Human values from digital activities.
\begin{axiom}
\label{axi:autonomousPrincipal} {\bf $a_i$ -Autonomous Principal} \\ There exist autonomous principals $a_i$ entities with an independent capacity to make choices. Autonomous principals are represented by a lower case latin letter. The index is ecosystem independent.
\end{axiom}
\begin{figure}[!h]
\center
    \includegraphics[width=40mm ]
    {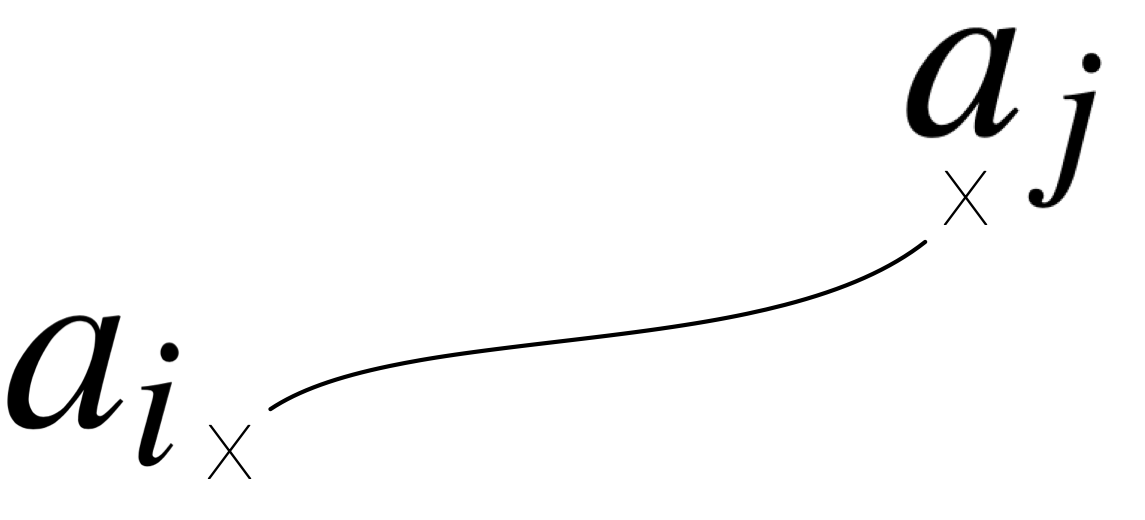}
    \caption{\textit{Connected autonomous principals} 
    \label{fig:autonomousAgentNormal}}
\end{figure}
This definition of autonomous principal based on free will excludes the engineering definition of autonomous principals whose decision making capacity is algorithmically based. Today's trust framework assign liabilities to autonomous principals, not algorithmic agents. Therefore, except a comment in section \ref{subsub:things} on page \pageref{subsub:things} and in the conclusion, the topics related to AI and IoT devices will be treated through the angle of data usage when dealing with data governance administration. 
\par
$a_i$ represent the autonomous Principal's {\it self} to which we can anchor both physical or digital identifiers. In most applications an $a_i$ will be part of multiple ecosystems and the same symbole has to be used across the different ecosystems. It is important to distinguish the symbole $a_i$ from the many identifiers to be used to identify $a_i$ in a given context. For example, $a_i$ can represent the physical  person opening a bank account or represent an account on a platform. The same $a_i$ has different representation in the physical and digital space but it refers to the same autonomous principal.   
\par
The governance model classify the autonomous principals according to their level of internal complexity. 
\begin{definition}{\bf Autonomous principals classification} \\
Autonomous principals $a_i$ can be assigned a {\sl type} defining classes of $a_i$'s according to the type of protection required. The core types defined below are immutably assigned to the autonomous principal, thus the type of principal is reflected by its symbol.
\end{definition}
We define three core classes of autonomous principals, humans, legal entities and sovereign entities. Specific use cases will augment and refine this typology. From a governance perspective, the types of autonomous principals are distinguished by the fundamental rights to be protected by the governance. 
\begin{subdefinition} {\bf $h_i$ Individuals}\\
Individuals are humans, protected by the right to privacy as recognised explicitly, for example, by article 8 of the European Convention of Human Rights \cite{ECHR:2021} or derived as the result of multiples torts as in the United States of America's {\sl "Right to be left alone"} \cite{Warren:1890,Prosser:1960};
\end{subdefinition}
\begin{subdefinition} {\bf  $o_i$ Organisations}\\
Organisations have a right of transactional sovereignty. Centralised control over the information impacts the freedom to transact both for public and private sectors. 
\end{subdefinition}
\par
If economic actors in the digital space do not enjoy the same level of protection provided by contract laws, trade secrets, confidentiality agreements or IP rights, unsustainable biases towards technology powerhouses are inevitable;
\begin{subdefinition} {\bf $g_i$ Political entities}\\
Politicial entities (e.g. countries) are protected by international laws to preserve their sovereignty \cite{Philpott:1995} centralised around a legitimate government recognised by other entities (e.g. an independent nation, provinces, states,) through international agreements.
\begin{figure}[!h]
\center
    \includegraphics[width=40mm ]
    {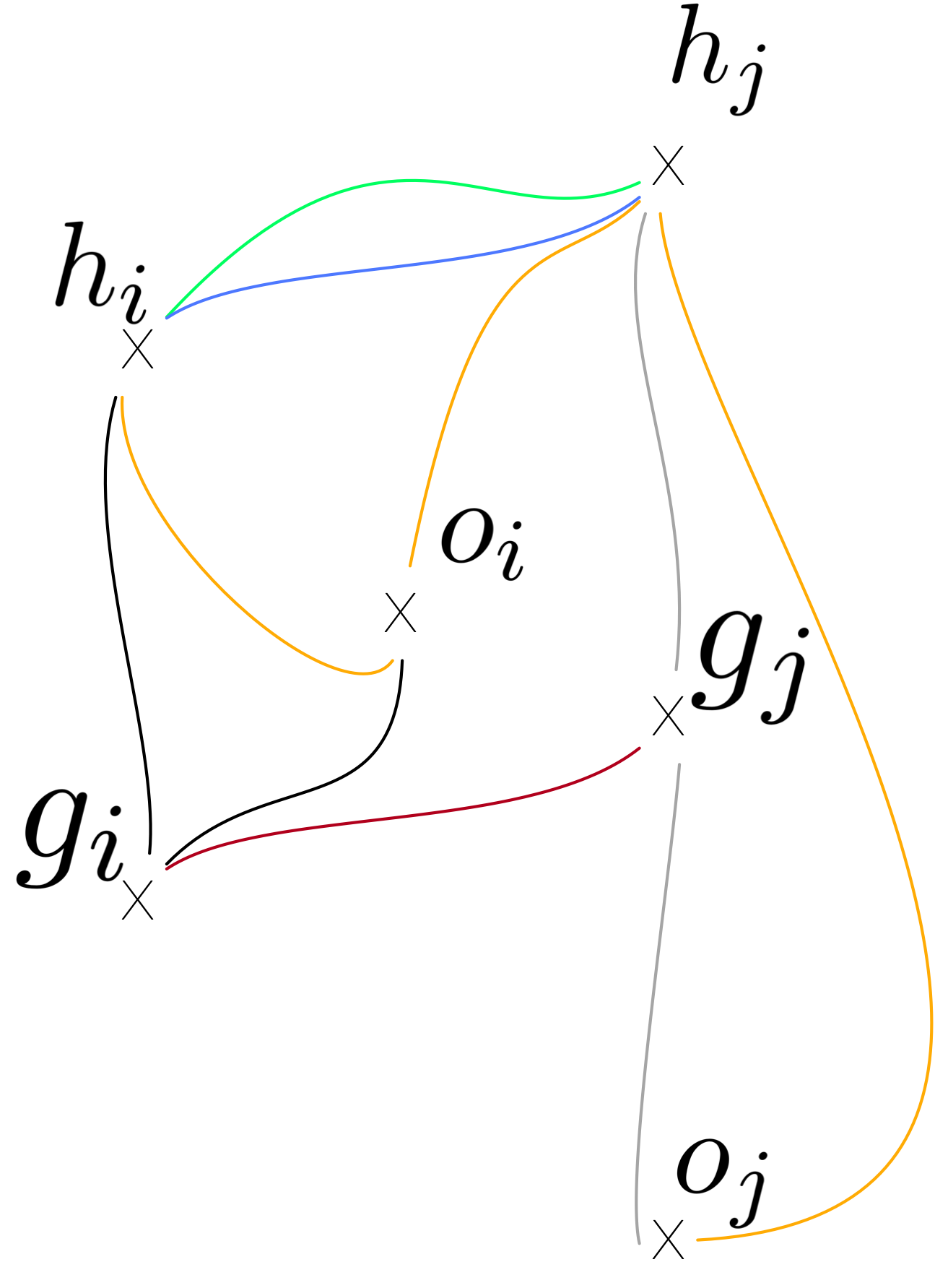}
    \caption{\textit{Multi-stakeholders \& Multi-jurisdiction network of connections. Colors indicates different governance regimes} 
    \label{fig:countryConnection}}
\end{figure}
\end{subdefinition}
\par
In principle, autonomous principals of any type can connect to any other type. The figure \ref{fig:countryConnection} represents the connections between citizens, organisations and public services.:
\par
These three core classes also represent levels different levels of complexity. The organising principle is the concept of {\sl ecosystem} which we introduce below.

\subsection{Ecosystems in Information networks\\{\sl Defining a dynamic units of governance}}
\label{subSec:ecosystems}
\subsubsection*{\sl Generalities} 
The distributed governance model is centred around the concept of {\sl ecosystem}. Simply stated, an ecosystem links together a group of autonomous principals into a governed environment. They are bound by a set of commonly agreed rules of conducts represented by a legitimate authority. Thus the defining elements necessary for a single ecosystem are simply its population\footnote{defined, in the statistical sense, as a e collection of principals under consideration for the ecosystem}, $\vec{a}_{\cal{A}}$, gathered around an authority $\cal{A}$ governing the ecosystem through an administration $\cal{A}^*$. The star notation represents the duality between the authority $\cal{A}$ and its its executive representative $\cal{A}^*$.
\par
In general, an autonomous principal can participate in many ecosystems, each defined with their own legitimate authority and administration. The governance of the system as a whole is determined by the trust relationships agreed between ecosystems  and enforced by their respective administrations. In the physical world, the unicity of the autonomous principal correlates its interactions with the different ecosystems. The autonomous principals interoperates between ecosystems based on the respective rules of each ecosystems as well as the mutual agreements between ecosystems, or adherence to multi-ecosystems treaties. 
\par
The distributed governance model is designed to integrate into a coherent framework the changes brought by digital transformation. Therefore, we define ecosystems independently of how the administration of the governance operates. The model only assumes the existence of rules governing the dynamics of the autonomous principals. 
\subsubsection*{\sl Re-think the Unit of Governance}
The novelty of the DGM approach is to recognise that a governance model requires unambiguous unicity of the autonomous principals and demonstrable causality in the chain of events leading to the accountability of a given autonomous principal. If obvious in the physical world made of matter evolving in space time, this is not the case in a digital space where both unicity and causality have to be constructed. Therefore these fundamental requirement to build the concept of {\sl unit of governance} are usually ignored. 
\par
Digital transactions live in the realm of information theory, not physical space. Therefore, these mechanisms have to be explicitly constructed and these are among the thorniest problems in computer science. As described in the introduction \ref{sec:introduction} information is spaceless (i.e. exists independent of its physical location) and timeless as the original singular unit (the bit) can not be distinguished from a copy\footnote{ the argument can also be extended to q-bits (quantum computing) but is outside the scope of this work}.
\par
Current system design are based on centralised solutions for authentication (i.e. no self-authentication) and time-stamping (i.e. no self-verifiable ordering of events) that prevent a functionally and verifiable independent unit of governance. The limitations of the current systems fuels the active development of decentralised authentication.
\par
The authors of this publication argue that decentralised authentication is necessary but not sufficient. Authentication must be extended from the authentication of actors to the decentralised authentication of passive objects. Therefore, the objectual integrity of data is a fundamental prerequisite to enable data oriented architecture. The necessity of objectual integrity lead to the development of decentralised semantic architectures. 
\par
\begin{figure}[!h]
\center
    \includegraphics[width=80mm ]
    {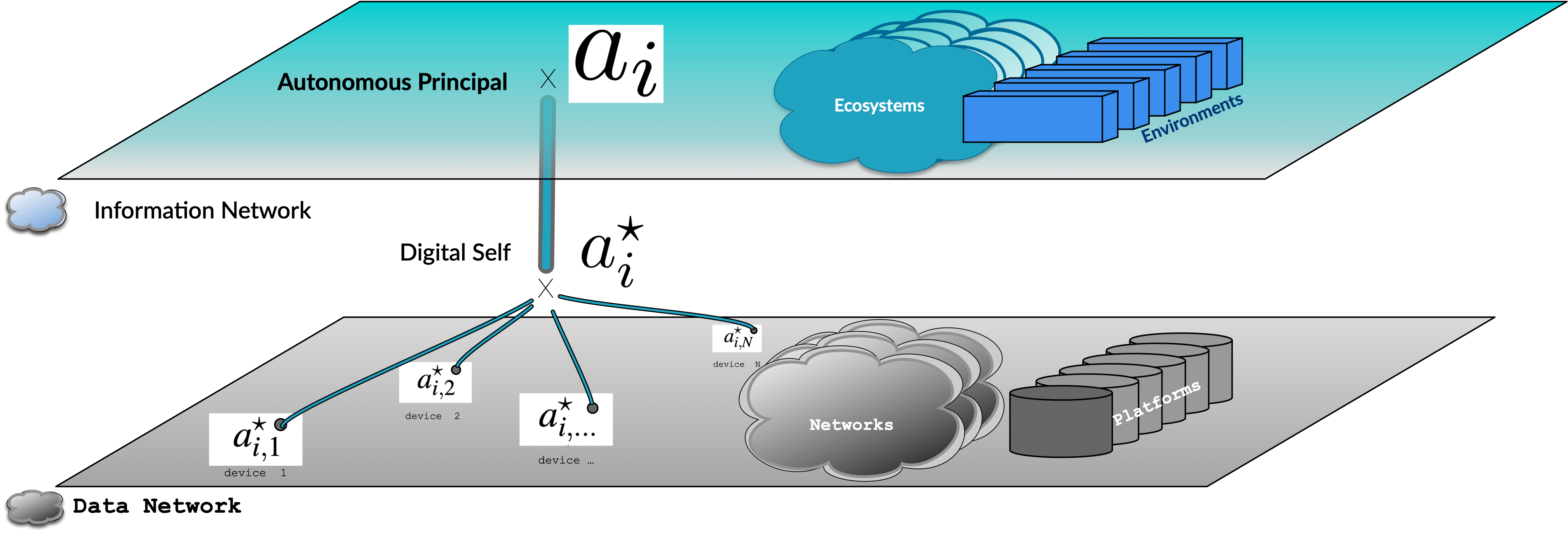}
    \caption{\textit{Information (Human) vs. Data (Technology) Network} 
    \label{fig:informationNetwork}}
\end{figure}

As a result, taken together, decentralised authentication and decentralised semantics introduce in the digital space the fundamental tooling upon which unicity and causality can be built without being locked into administrative technological intermediaries. DCM is designed to provide the contextual and ecosystem driven governance with the assumption that the unit of governance, the autonomous principal $a_i$ can be bound to its digital representation $a_i^\star$ (as illustrated in figure \ref{fig:informationNetwork}).
\par
Section \ref{sec:consensualVeracity} provides an instance of a data oriented architecture coined {\sl Dynamic Data Economy}  integrating decentralised authentication, decentralised semantics and distributed governance. 
\par
Recognising the above explains the central role of a functional definition of ecosystem to match the complexities of the intricacies of Human rules and regulations.
\begin{definition} {\bf $\rm{E}_{\cal{A}}$ Ecosystem} \\
  An ecosystem is defined as a self-contained unit of governance. It is defined by a triplet consisting of the legitimate authority, the population of autonomous principals adhering to this authority  and the administration enforcing the ecosystem governance:
\label{def:ecosystem}
\begin{equation}
{\rm E}_{\cal{A}} =  \left(
	\vec{a}_{\cal{A}}
	, \cal{A}, \cal{A}^*
	\right) 
\end{equation}
 \end{definition}
In specific models,  $\rm{E}_{\cal{A}}$, notations will simplifies to $\cal{A}$ as autonomous principals often refers the ecosystem in reference to the legitimate authotity (e.g. I am a citizen of $\cal{A}$).
\par
The population $\vec{a}_{\cal{A}}$ of an ecosystem is defined by the set of autonomous principals subject to the governance of the legitimate authority. The generic notation might appear cumbersome but it reflects the fact that the autonomous principal $a_i$ exist independently of the ecosystem as well as the link between different ecosystems. It will simplify and proves useful when applied to specific use cases.
\begin{definition} {\bf $\vec{a}_{\cal{A}}$ Population}\\
\label{def:ecosystemPopulation}
\begin{equation}
\begin{array}{lcl}
\vec{a}_{\cal{A}} & = & 
	\left\{
		a_i\vert_1^{\cal{A}}, a_j\vert_2^{\cal{A}}, \ldots, a_k\vert_{N( I_{\cal{A}})}^{\cal{A}}
	\right\} \\
I_{\cal{A}} & = &
	(
		 1, 2, \ldots , N( I_{\cal{A}})
	) \\
	& = & {\rm ecosystem's\; population\; indexing}\\
\end{array}
\end{equation}
$N( I_{\cal{A}})\;=\;  {\rm Card}( I_{\cal{A}} )$ is the population size.
The population of an ecosystem is the set of autonomous principals that recognise to be subject to the legitimate authority $\cal{A}$. In other words, the principal consent in principle to the ruling of  $\cal{A}^\star$. The application of a specific rule depends on the context of a given transaction. For example, as a citizen of $\rm{E}_{\cal{A}}$ traveling in country $\rm{E}_{\cal{B}}$, you implicitly consent to be subject to the subset of $\cal{B^\star}$ applied to foreigners.
\end{definition}
The main role of the legitimate authority is to represent the binding of the population $\vec{a}_{\cal{A}}$ gathered around $\rm{E}_{\cal{A}}$ . To achieve this, the legitimate authority must hold both representatives of the authority and the reference documents upon which the authority realises to legitimate the action of its governance administration.  $\cal{A^\star}$ defines the rules and regulations to be obeyed by the ecosystem's population. For certain use cases (e.g. sovereign entity) the legitimate authority is represented by one or more autonomous principals. As  introduced in the previous section, these governing principals are differentiated and represented by $g_i$. 
\begin{definition} {\bf ${\cal A}$ Legitimate Authority}
\label{def:ecosystemLegitimateAuthority}
\begin{equation}
\begin{array}{lcl}
{\cal A} &=&
	\left\{ g_i \right\}_{i\in G_{\cal{A}}} 
	\bigcup
	\left\{ R_i \right\}_{i\in \cal{R}_{\cal{A}}} 
	\\
\left\{ g_i \right\}_{i\in G_{\cal{A}}} &=& {\rm set\; of\; authority \;representatives} \\
\left\{ R_i \right\}_{i\in \cal{R}_{\cal{A}}} &=& {\rm set\; of\; authoritative \; documents}
\end{array}
\end{equation}
\end{definition}
${\cal A}$ is the underlying authority defining the ecosystem and providing the legitimacy of the ecosystem's governance. As such $\cal{A}$ must have at least a set of autonomous principals $g_i$ representing the authority as well as a set of reference documents trusted by the ecosystem's population.  Generally the autonomous principals $g_i$ do not interact directly with the ecosystem constituants $\vec{a_i}$. It is the role of a set of separate organisation to enforce these rules $\cal{A}^\star$, the governance administration.
The governance administration is the tangible representation of the legitimate authority. Therefore, its structure parallels the one of the authority itself.
\begin{definition}{\bf ${\cal A}^\star$ Governance Administration}\\ 
\label{def:ecosystemGovernanceAdministration}
\begin{equation}
\begin{array}{lcl}
\cal{A}^\star &=&
	\left\{
		\cal{A}^\star_{\sigma\in G^\star_{\cal{A}}} 
	\right\}
	\bigcup
	\left\{
		\cal{R}^\star_{\sigma\in R^\star_{\cal{A}}} 
	\right\}
	\\
\left\{
	\cal{A}^\star_{\sigma\in G^\star_{\cal{A}}} 
\right\} & = &
	{\rm set\; of\; administrative \; entities}
	\\
\left\{
	\cal{R}^\star_{\sigma\in R^\star_{\cal{A}}} 
\right\} & = & 
	{\rm set\; of\; govering \; references}
	\\
\end{array}
\end{equation}
$\cal{A}^\star_{\sigma\in G^\star_{\cal{A}}} $ and $\cal{R}^\star_{\sigma\in R^\star_{\cal{A}}} $ are the services required to  establish and enforce the ecosystem's governance. They enable  $\rm{E}_{\cal{A}}$ to exist as a cooperative entity serving its population and protecting the community. This includes the rules and regulations internal to the ecosystem but also the cross-jurisdictional rules for interactions across ecosystems $\rm{E}_{\cal{A}}$ and $\rm{E}_{\cal{B}}$
\par
The heavy notation for the abstract distributed governance model is required to encompass the wide variety of use cases where the interplay between traditional (human) governances and technologically driven digital governances requires novel approaches.
\par The aim of this series on Dynamic Data Economy is to provide an explicit construction of information governance administration for specific use case. Thus an explicit form of a governance administration for information systems in term of registries, consent management, data exchange is reserved for part 4 once the enabling technologies have been detailed.
\end{definition}
\input{table-cv2-notationAbstractGov.tex}

\subsubsection*{\sl Ecosystem complexity levels}
\begin{figure}[!h]
\center
    \includegraphics[width=80mm ]
    {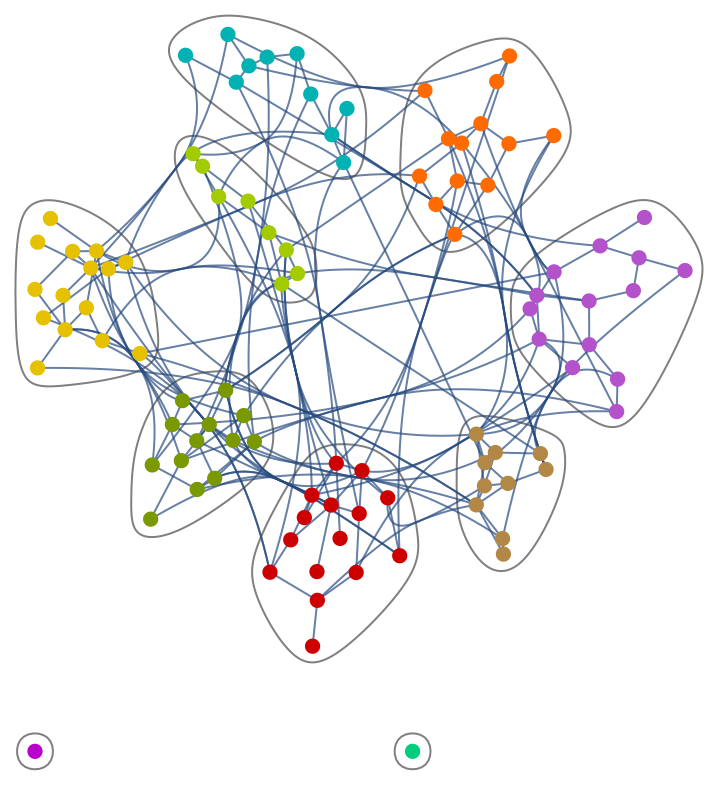}
    \caption{\textit{Graph representation of interacting ecosystems} 
    \label{fig:communityGraph}}
\end{figure}

Looking at the ecosystem $\rm{E}_{\cal{A}}$ as a "unit of governance" demands that we consider both the internal governance of the ecosystem (i.e. how the ecosystem establishes and enforces its own rules) and the external governance (i.e. the rules dictating information exchange between ecosystems).  This distinction can be achieved simply by considering the ecosystem itself as an autonomous principal when $\rm{E}_{\cal{A}}$ acts with its peers.  For example, exchanges between 10 sovereign countries are regulated by bilateral agreements or adherence to international treaties as well as by movements of individuals between countries as represented in figure \ref{fig:communityGraph} on page \pageref{fig:communityGraph}. The color coding refers to the separate governance of each countries. This example includes two communities not interacting with the others.
\par
With this goal in mind, the definition the autonomous principal $a_i$ itself as the lowest level of complexity for an ecosystem allows an iterative construction of ecosystems of increasing complexity. This lowest level ecosystem is necessary to build a consistent model.
\begin{definition}{\bf $\rm{E}_{a_i}$ -Lowest Level Ecosystem}\\
\label{def:lle}
The lowest level ecosystem is defined as the autonomous principal itself:
\begin{equation}
\begin{array}{rcl}
E^0_{a_i} 	& = & \left(
				\left\{ a_i  \right\}
				,
				a_i
				,
				a_i
			\right) \\
& \equiv & a_i \\
with &&\\
I^0_{a_i}  & = &  \left\{ 1 \right\} \\
N( I^0_{a_i}) & = & 1 
\end{array}
\end{equation}
\end{definition}
This definition stresses the fact that the distributed governance model is fundamentally based on peer-to-peer interactions even if the autonomous principal is itself a complex entity with an internal structure. This provides the model with a property upon which a dynamic of evolving ecosystems can be build. 
\par An ecosystem has a liberty of action defined by its authority, usually in co-ordination with other ecosystems. Thus an organisation or a political entity also has the capacity of choice and will therefore be accountable for its decisions. From this perspective, an ecosystem is also an autonomous principal with the capacity to create or join other ecosystems. Ecosystems have a spontaneous-replicating capacity. An ecosystem is therefore an autonomous principal of higher complexity. A society becomes a mesh of ecosystems interacting under the drive of autonomous principals. The spontaneous-replicating properties for an ecosystem  emerges from the capacity of autonomous principals to make choices \ref{axi:autonomousPrincipal}. Therefore we present it as an axiom:
\begin{axiom}{\bf spontaneous-replication}\\
\label{axi:selfReplicating}
An ecosystem $\rm{E}_{a_i}$ has the capacity to create new ecosystems $\rm{E}_{b_j}$ by itself or in cooperation with other ecosystems.
\begin{equation*}
\rm{E}_{a_i}, \ldots, \rm{E}_{a_k} \longmapsto \rm{E}_{b_j}
\end{equation*}
\end{axiom}
\subsubsection*{\sl Dealing with {\sl Things} }\label{subsub:things}
The governance of Artificial Intelligence (AI) and automated devices (IoT) is a key unresolved problem. As a result, we face both societal problems and a slow down the digital transformation. One of the aims of the distributed governance model is to provide a framework upon which a coherent governance embedding these technologies can be designed. Our definition of autonomous principal encompassing both physical and moral persons (ecosystems), provide a framework in which {\sl Things} can be defined.
\par
We consider  {\sl Things} as non-autonomous entities defined in opposition to autonomous entities $a_i$.
\begin{axiom} {\bf $\sigma_\alpha$ -Non-autonomous agent}\\
\label{axi:nonAutonomousAgent}
There exist non-autonomous agents $\sigma_\alpha$ representing devices or systems whose actions are algorithmically (or mechanically) driven and ultimately under the accountability of an autonomous principal $a_i$. Non-autonomous entities are labeled with lower case greek letters from the end of the alphabet.
\end{axiom}
In other words, things do not have the capacity of accountable choice. For example, a self-driving car has both an owner and a constructor. In the debate of the responsibility in an accident, ultimately the community (i.e. the ecosystem) will decide who is accountable and design the governance around it. And these decisions will be dependent of the jurisdiction. In the case of weapons with automated lethal capacity, it is already clear that the sovereign entity commissioning the usage of the weapon is accountable. The governance (e.g. ban) of such weapon is debated within international forums, on humanitarian ground, not technology.
\par
In today's world, systems whose actions are the result of mechanical or algorithmic decisions fits this definition of {\sl Things}. They are an essential part of networks and provide a quantum leap in the way decisions are made by providing near-realtime factual information.
\par Specifically, from a distributed governance, {\sl Things} are passive objects ultimately under the control of an autonomous principal $a_i$ or a governance authority ${\cal A}^\star$ of an ecosystem. Thus they will formally appear in the explicit construction of governance administration and do not need a more detailed definition at this stage. 
\par
A more in-depth treatment of non-autonomous entities and their interactions with autonomous principals $a_i$ is treated in a subsequent publication.
%
\subsection{Extrinsinc \& Intrinsic properties of data\\ {\sl Human trust vs. Technological assurance}} 
\label{subSec:extrinsic}

Data protection regulations protect individuals, not data. Professionals in this field, like data protection officers, know how to navigate between legal, societal (business) and technology requirements. They understand the complementary role of data protection and cybersecurity. However, in practice, the intricate nature of the two domains leads to complexities that are hard to understand for users or decision-makers impacted by digital transformation. 
The distributed governance model helps to clarify technology's role in human governance with a clear distinction between intrinsic and extrinsic properties of data.
\par
For example, the mass of a physical object is an intrinsic property, while its weight is an extrinsic property; it depends on an external factor, the gravitational field in which the object evolves. With the same mass, an object has a different weight on earth or the moon. In the model, we want to apply these definitions to information theory in analogy with the physical world. 
\par
The extrinsic vs intrinsic analogy in the data space is made possible with the recent development of decentralised semantics and decentralised authentication applied to passive identifiers. The concept of objectual integrity introduced in section \ref{subsec:objectualIntegrity} on page \pageref{subsec:objectualIntegrity} leads to data objects with verifiable integrity. The object's definition is assessed without needing an external intermediate party. Then factual authenticity (\ref{subsec:factualAuthenticity} on page \pageref{subsec:factualAuthenticity}) can add a cryptographic unicity of the object and an immutable causality of its digital evolution.
\par
As a result, a digital object can acquire an independent digital existence with the same essential properties as physical objects through their unicity and ordering of events affecting them. Therefore, these properties provide the basis for distinguishing data's extrinsic and intrinsic properties.
\par
In this work, we use the following definitions. 
\begin{eqnarray}
{\rm data} & \equiv  & {\rm intrinsic\;properties}\\
{\rm information} & \equiv & \begin{array}{c}
{\rm data} \\
+\; {\rm extrinsic\; properties}
\end{array}
\end{eqnarray}
\begin{definition} {\bf Data extrinsic properties}\\
\label{defi:extrinsicProperties}
An ecosystem's governance and its economic domain define data {\sl extrinsic properties}. They assign meaning to data objects upon which data acquire a value or a specific role. The extrinsic properties of data link human values and community definitions to a representation of the information we define as data. For example, private, confidential, legal, and monetary properties are defined by society and used by autonomous principals. 
\\
Extrinsic properties of data allow the functional definition of societal concepts like privacy, identity and reputation.
\end{definition}
For example, the cryptographic properties of a digital token are intrinsic, while the token's value is an extrinsic property. Similarly, the definition of an {\sl attribute type} is an intrinsic property of a data object while its qualification as personally identifiable information (PII) is extrinsic. 
\par
\begin{figure}[t]
    \includegraphics[width=\linewidth,
   ] {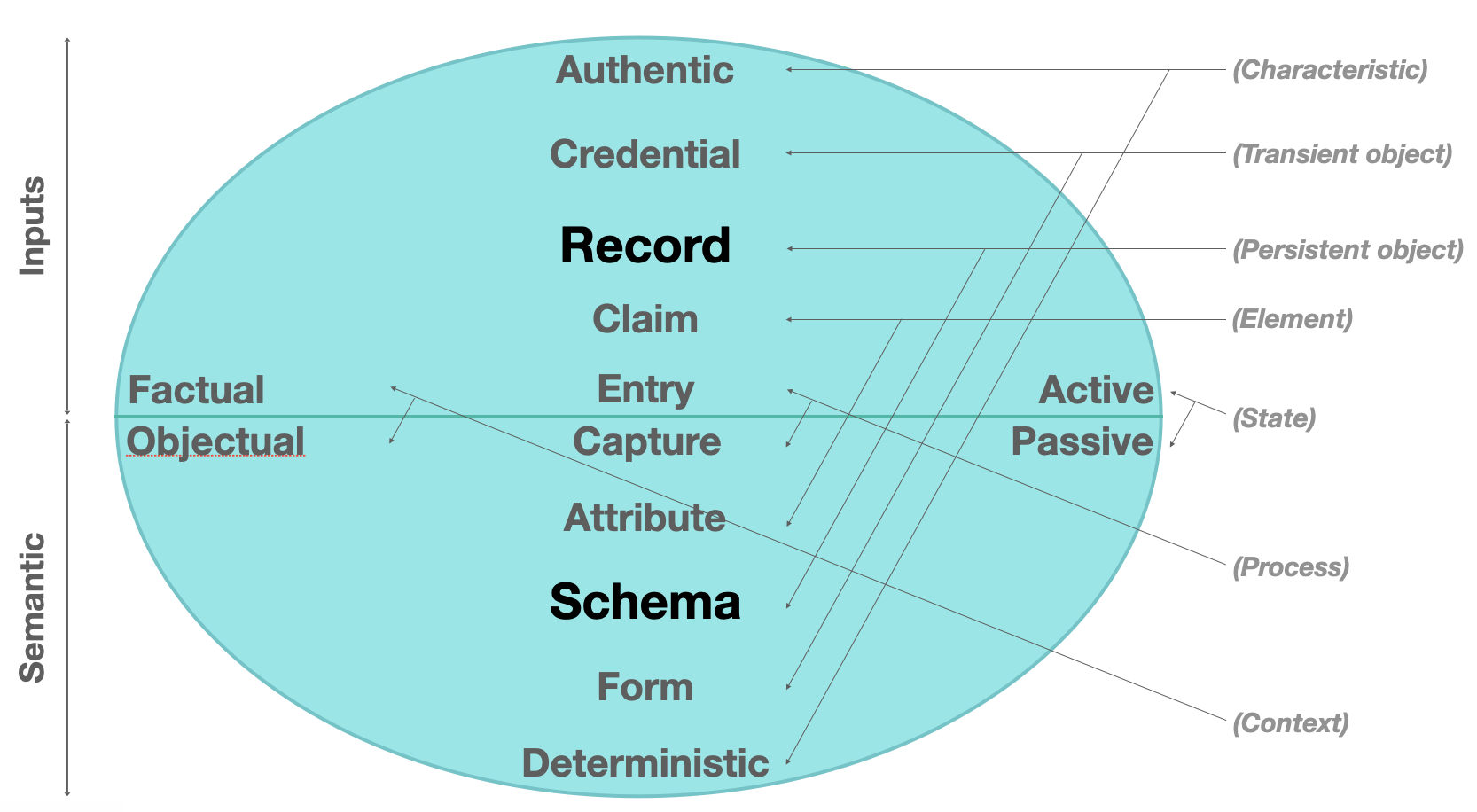}
    \caption{Intrinsic properties within DDE's {\sl Data Network Model}}
    \label{fig:dataNetworkModel}
\end{figure}

\begin{definition} {\bf Data intrinsic properties}\\
\label{defi:intrinsicProperties}
The {\sl intrinsic properties} of a data object are defined by the data models They are intimately linked to a representation, not necessarily digital. They are independent of their actual usage. For digital representation, as data are subject to algorithmic transformation, a wealth of internal properties have emerged. A network model allows a classification that is necessary for proper description of data usage.
\end{definition}
In section \ref{sec:consensualVeracity} we introduce the Dynamic Data Economy architecture based on a {\sl Rugby Ball} data network model represented in figure \ref{fig:dataNetworkModel}. The model separates the data capture part, the process of collecting data electronically, allowing it to be stored, searched, or organised more efficiently (i.e. the {\sl Inputs Domain}) from the data entry in the {\sl Semantic Domain}\cite{HCF:2022_06}.  These subjects are treated in the second and third part of this series. For sake of completeness, we simply list here the different items and introduce the notation.
\begin{definition}{\bf $\cal{M}$ -Data Network Model}
The physical structure and flow of data between participating entities. The data Network secures the intrinsic properties of data.
\end{definition}

\setcounter{subdefinition}{0}
\begin{subdefinition} Inputs Domain\\
What is put in, taken in, or operated on by any process or system.
\end{subdefinition}
\begin{subdefinition} Semantics Domain\\
The meaning and use of what is put in, taken in, or operated on by any process or system.
\end{subdefinition}
\begin{subdefinition} $\cal{R}$ Context \\
the circumstances that form the setting for an event can be
		\begin{enumerate}
			\item {\sl Factual}: Using or consisting of facts.
			\item {\sl Objectual}: Relating to or represented as an object.
		\end{enumerate}
\end{subdefinition}
\begin{subdefinition} $\cal{P}$ Process \\
a series of actions or steps taken in order to achieve a particular purpose with data:
		\begin{enumerate}
			\item {\sl Data Entry}: inputting data into a computer using devices such as a keyboard, scanner, disk, sensor, or voice
			\item {\sl Data Capture}: collecting data electronically, allowing it to be stored, searched, or organised more efficiently
		\end{enumerate}
\end{subdefinition}
\begin{subdefinition} $\cal{S}$ State \\
 the particular condition of an identifier
		\begin{enumerate}
			\item {\sl Active}: A type of identifier that requires a signing key to authenticate and bind an active governing entity to an event.
			\item {\sl Passive}: A type of identifier that has an association with a cryptographic hash of digital content, which acts as a deterministic fingerprint to identify passive objects and their relationships.
		\end{enumerate}
\end{subdefinition}
\begin{subdefinition} $\cal{C}$ Characteristics\\
 {\sl typical of a particular data}
		\begin{enumerate}
			\item {\sl Authentic}: True, quantitative or qualitative information collected from real-life phenomena. Data can be assumed to be authentic if it is provable that it has remained incorrupt since its creation.
			\item {\sl Deterministic}: No randomness is involved in the development of future states of the object. If any operation's result and final state depend solely on the initial state and the operation's arguments, the corresponding object is deterministic.
		\end{enumerate}
\end{subdefinition}
\begin{subdefinition} ${\cal O}_T$ Transient Object \\
 {\sl for impermanent data}
		\begin{enumerate}
			\item {\sl Credential}: A piece of any transient document detailing a qualification, competence, or authority issued to an individual by a third party with a relevant or de facto authority or assumed competence to do so.
			\item {\sl Form}: The digital equivalent of a transient paper document used to capture data.
		\end{enumerate}
\end{subdefinition}
\begin{subdefinition} ${\cal O}_P$ Persistent Object \\
{\sl for potentially permanent data}
		\begin{enumerate}
			\item {\sl Record}: A collection of data items arranged for processing by a program. Records in a persistent database or spreadsheet are "rows", a collection of fields, possibly of different data types, typically in a fixed number and sequence.
			\item {\sl Schema}: The organisation or structure for a persistent database, while in artificial intelligence (AI), a schema is a formal expression of an inference rule.
		\end{enumerate}
\end{subdefinition}
\begin{subdefinition} $\cal{E}$ Element\\
{\sl essiential parts of a data object}
		\begin{enumerate}
			\item {\sl Claim}: A piece of packaged information signed into a security token and stored.
			\item {\sl Attribute}: A single-value descriptor for a data point or data object.
		\end{enumerate}
\end{subdefinition}
The governance administration of an ecosystem governs the dynamics of the extrinsic data properties to secure the information exchanges driven by autonomous principals. For example, the properties leads to the definition of assurance levels that  the technology infrastructure implements via standards (or proprietary solutions) for handling internal properties of data. 
\par
With the above definitions the information lives in a space with dual components an {\sl Information Network} intimately linked to, but distinct from the Data Network.
\par
{\sl A note on digital identity}\\
 This distinction is particularly helpful when dealing with situations involving {\sl Identity}. The same term is used to represent both an extrinsic (notional identity) and intrinsic (authentication) property of the information. Reducing the concept of identity to authentication leads to drastic simplification most communities will not accept. This is more clearly seen once a strong link between the autonomous principal and the identifiers under his control is established.
\par
The following subsection \ref{subSec:confidentialitySpheres} explains how data objects are unambiguously related to a controlling agent. When this controlling agent is an autonomous principal $a_i$, the ecosystem links the principal to the applicable governance protecting his rights and defining its responsibilities.
\par
\input{table_networkCharacteristic.tex}

\subsection{Confidentiality Spheres\\{\sl Planting the seeds of consent}}
\label{subSec:confidentialitySpheres}
The central element upon which a governance is anchored, the autonomous principal $a_i$ (sec. \ref{subSec:autonomousPrincipals}) while the governance itself is defined by ecosystems $\rm{E}_{\cal{A}}$ (sec.\ref{subSec:ecosystems}) to protect the community it represents. This section defines how the principals interacts among themselves.
\par
We introduce the concept of {\sl confidentiality spheres} as a critical element to be present at the core of any information exchanges. The autonomous principal is aware of the context in which the information exchange takes place and therefore can asses the level of privacy to be imposed in function of the trust in the other party and its reputation.
\par
When exchanging information, cooperation and trust can only be enabled if the confidentiality level of the communication is a core intrinsic property of the exchange. In the physical world, it is a natural and unconscious reflex for individuals and organisations alike. Individuals constantly re-asses the context and adjust the communication content to the level of trust in the receiving party. Organisations have developed similar risk mitigating measures with information exchange covered by contractual agreement (e.g. non-disclosure agreements, power of attorneys) and regulatory protection.

\subsubsection*{Privacy Spheres} Originating in privacy laws, the concept of privacy sphere \cite{Simitis:1982,Mokrosinska:2018,Page:1982} came forward to protect individual freedoms. As an intangible legal object derived from other laws (e.g. swiss civil code art.28), it nevertheless helps to define the framework of laws and regulations around privacy. A privacy sphere is a construct protecting the individual (the autonomous principal) from the environment by defining what are private information; information not accessible without explicit consent of the individual or a legitimate authority. In the physical world, privacy is protecting freedom through different laws and regulations pertaining to a specific ecosystem. In information systems, this concept can be used to provide a functional definition of privacy to be implemented with tools like Personally Identifiable Information (PII) or Privacy Enhancing Technologies (PET). The protection offered by PII and PET is limited when they are implemented at the application level. 
\par
The distributed governance model builds upon the privacy sphere concept to define {\sl confidentiality levels} as a core design concept. First when applied to information systems, the functional construct can be extended outside privacy laws to private law\cite[p.24]{Page:1982} and also allows an economic approach\cite[p.125-146]{Page:1982}. Second it is not bound to a specific jurisdiction. Multi-jurisdiction solutions to be developed for international governance \cite[p.165,p.69 ]{Page:1982} or domain specific jurisdictions \cite[p. 171]{Page:1982}.  As a result, the privacy sphere concept, when functionally defined at the autonomous principals  interaction level, applies in both physical and digital spaces.

\subsubsection*{Confidentiality levels} 
Following the privacy sphere model, we formalise three confidentiality levels based on the existing legal constructs of public, private, and intimate sphere.
\begin{description}
\item {\bf Intimate} for information exchanged with known trusted partners, usually a few close receiving agents well known by the sender. This connection should imply reciprocity. The patient-doctor relationship is an example of a contextual. intimate relationship that is actually enforced by laws, best-practices, and ethics. The connection type can also be used between an autonomous principal and a passive device controlled by the principal to ensure a strong human-hardware binding.
\item {\bf  Private} for information exchanged with a closed community. The perimeter is a subset of members of an ecosystem that can be defined and made transparent to the sender if needed. For an organisation this is often contractually enforced by non-disclosure agreements for example. For individuals it can be defined by specific legal instruments like data protection laws (i.e. GDPR) or contextually agreed definitions like PII (Personally Identifiable Information) in anglo-saxon jurisdictions.
\item {\bf Public} for information exposed to an unknown, possibly unlimited audience. The sending agent implicitly consent that he loses control of the information made public. Once in the public domain, the information relies solely on the community rules to protect him from information misuse. The anti-diffamation regulations are a good example of a governance set forth by the community to protect its member. In this specific example it is clear that the level of tolerance are highly dependent of the community itself. It is important to notice that the rules set by a community also protect the receiver agent from information willingly made public by a sending agent. "Hate speech" regulations are a good example. Similarly for organisation, public information is protected by ecosystem dependent regulations (e.g. IP rights).
\end{description}
{\sl\bf Privacy Sphere for autonomous principals} 
\par
This work is intended to address governance issues resulting from digital transformation. Therefore, we provide a set of definitions describing the communication framework between autonomous principals that can be used both with digital and non-digital means. In this framework, the privacy sphere concept can receive a functional definition that can be applied to wide variety of cases. Privacy spheres are defined as a set of autonomous principals connected together. The confidence levels of the connection is defined in terms of a property the sender assigns to either a connection or the messages within.
\par
In this section we do not address the communication between autonomous principals and {\sl things} (i.e.non-autonomous agent). Autonomous principals can be liable of their decisions while {\sl things} comply to assurance levels only.
\\
{\sl\bf Messages \& Connections}
\par
 A {\sl message} \ref{def:messages} is any signal send by one entity (the sender) to a different entity (the receiver). Both the sender and the receiver might have a structure. For example, the sender can be an individual (e.g. a citizen) communicating with an organisation (e.g. administrative service). The definition only assumes senders and receivers can be identified and, in the case of autonomous principals as the controller of the sender identifier, also controls the signal emission.  We differentiate between {\sl active message} and {\sl passive message} to distinguish different level of control of the sender. 
\par
A message has a direction and duration. It starts when emitted by the sender and ends when received by the receiver. Messages are transient by nature. They differ from {\sl connections} \ref{def:connection} that, even if inactive, are permanent until terminated. 
\begin{definition}[$\ldots > \dots$ Messages]
\label{def:messages}  $\forall s_i, r_j \in \cal{W}$, let
\begin{eqnarray*}
s_i & = & {\rm sender} \\
r_j & = & {\rm receiver}
\end{eqnarray*}
a message is any signal $>$ send by $s_i$ to $r_j$ within $\cal W$. A generic message is written as:
$$
s_i \quad > \quad r_j
$$
\end{definition}
\begin{subDefinition}[active ($\longrightarrow$) vs. passive ($\longleftrightarrow$) message]
\label{def:activeMessage} \label{def:passiveMessage}
If the message emission is under the direct control of $s_i$ it is defined as {\sl active}. An active message always has an intended receiver. If the message emission is under the control of the context in which $s_i$ is present, it is defined as {\sl passive}. In such case, the receiver might not be identifiable. 
Active vs. passive messages are represented by:
\begin{eqnarray*}
s_i > \longrightarrow   r_j & {\rm active \quad message} & \\
s_i > \longleftrightarrow \ldots & {\rm passive \quad message} &\\
\end{eqnarray*}
\end{subDefinition}
A {\sl connection} is a set of messages within a context and between two entities. Connections have no natural duration, once created, might remain open indefinitely. Connections have no natural direction, both entities are connected but remain independent.
\par
For the case of autonomous principals $a_i$:
\begin{definition}[$C_{ij}$ -Connections]  are a set of messages $>$ between $a_i$ and $a_j$.
\begin{eqnarray}
\label{def:connection}
C_{ij} & = & \lbrace a_i \quad\underrightarrow{>_\alpha}\quad a_j  \mid \alpha \in \Lambda \rbrace
 \bigcup
 \lbrace a_j \quad\underrightarrow{>_\beta}\quad a_i  \mid \beta \in \Lambda \rbrace 
\end{eqnarray}
where $\Lambda$ represents the set of messages in a given confidentiality level.
\end{definition}
The confidentiality levels are a property that can be assigned to messages.
\begin{definition}[Confidentiality levels]
\label{def:confidentialityLevels}  The confidentiality level is an intrinsic property of the connection between entities and messages within a connection. For a connection, the confidentiality level is unique and equally controlled by participating entities and/or the context in which the connection takes place. Within a connection, messages might have different confidentiality levels. The sender controls the confidentiality level (i.e. not the receiver). 
\end{definition}
\begin{subDefinition}[$\unrhd$ -Intimate message]
\label{def:intimateMessage} 
The receiver of an intimate message is expected to be known even if the message is passive. The message content or the context inform the receiver of the confidentiality level.
Intimate messages are represented by:
\begin{eqnarray*}
 s_i \unrhd  \longrightarrow  & \ldots & r_j\\
\end{eqnarray*}
\end{subDefinition}
\begin{subDefinition}[$\rhd$ -Private message]
\label{def:privateMessage} 
The message content or the context inform the receiver of the private confidentiality level.
Private messages are represented by
\begin{eqnarray*}
 s_i \rhd \longrightarrow  & \ldots & \\
\end{eqnarray*}
\end{subDefinition}
\begin{subDefinition}[$\gg$ -Public message]
\label{def:publicMessage} 
The message content or the context inform the receiver of the public confidentiality level.
Public messages are represented by
\begin{eqnarray*}
 s_i \gg  \longrightarrow  & \ldots & \\
\end{eqnarray*}\end{subDefinition}
The table below provide a few examples. 
\input{table_exampleMessage.tex}

In diagrams, messages are represented by straight lines.
\begin{figure}[!h]
\center
    \includegraphics[width=40mm ]
    {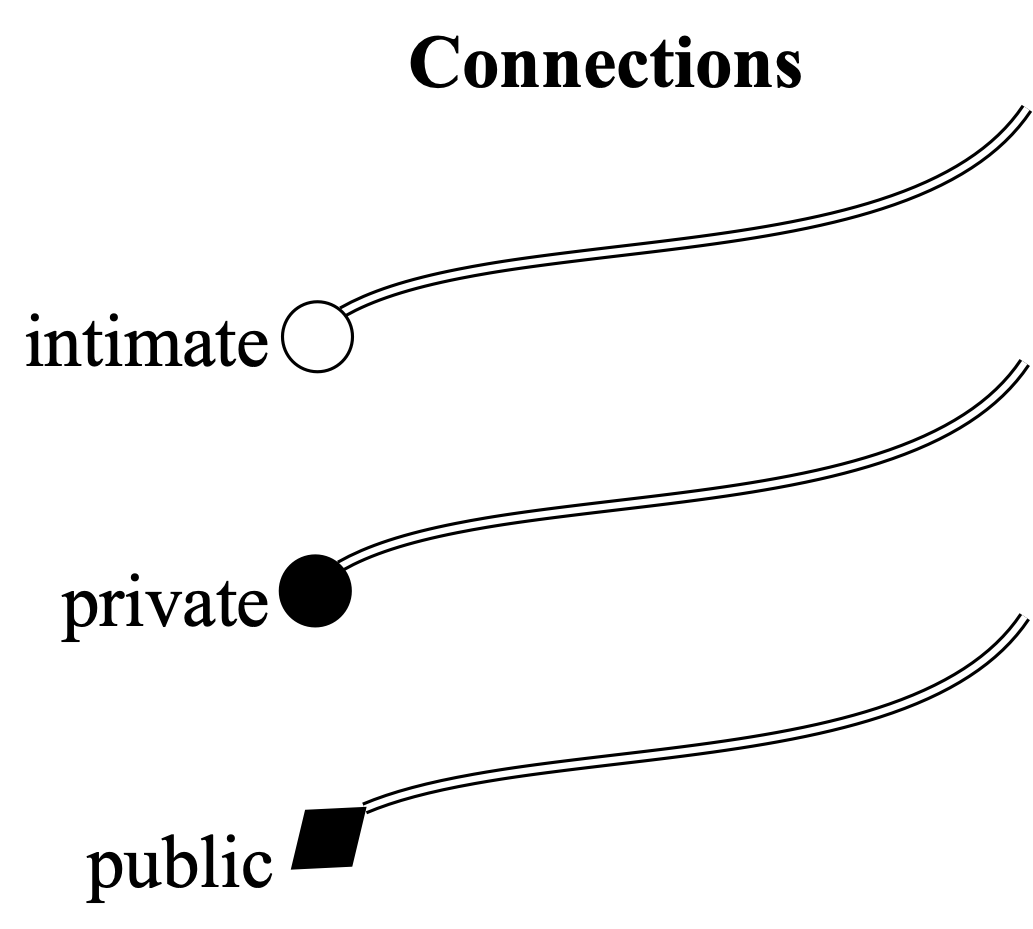}
    \caption{\textit{Messages representations} 
    \label{fig:messages}}
\end{figure}
The formalism for intimate (\ref{def:intimateMessage}), private (\ref{def:privateMessage}) and, public (\ref{def:publicMessage}) messages can be extended to connections :
\begin{eqnarray}
C_{i\unrhd j} & = & {\rm an\;intimate\;connection}\\
C_{i\rhd j} & = & {\rm a\;private\;connection}\\
C_{i\gg j} & = & {\rm a\;public\;connection}
\end{eqnarray}
It is important to note that the confidentiality level of messages can be different than the ones of the connection. More precisely, one could expect that the following definitions could be used:
\begin{eqnarray*}
C_{i\unrhd j} & = & \lbrace a_i \quad\underrightarrow{\unrhd_\alpha}\quad a_j  \mid \alpha \in \Lambda_\unrhd \rbrace\\
C_{i\rhd j} & = & \lbrace a_i \quad\underrightarrow{\rhd_\alpha}\quad a_j  \mid \alpha \in \Lambda_\rhd \rbrace\\
C_{i\gg j} & = & \lbrace a_i \quad\underrightarrow{\gg_\alpha}\quad a_j  \mid \alpha \in \Lambda_\gg \rbrace
\end{eqnarray*}
However, the above are not generic and only represent a certain "pure" use cases. A private connection might contain public messages or intimate messages when information is to be kept ”off the record" for example.

\subsubsection*{\sl Confidentiality Sphere}
\begin{definition}[$\mathcal{C}_{a_i}$ -Confidentiality Sphere] of an autonomous principal $a_i$: is the set of autonomous principals $a_k$ interacting with $a_i$ through connections of a given confidentiality level. $a_i$ controls $\mathcal{C}_{a_i}$. The degree of control is function of the confidentiality level of $\mathcal{C}_{a_i}$. 
\begin{eqnarray}
\mathcal{C}_{a_i} & = & \lbrace a_j  \bigcup C_{ik} \mid \forall \; k \; for \;which\; \exists\; C_{ik} \rbrace
\end{eqnarray}
where $C_{ik}$ are the connections defined by equation \ref{def:connection} on page \pageref{def:connection}.
\end{definition}
leading to the natural definition of intimate, private and public sphere of the agent $a_i$.
\begin{eqnarray}
\mathcal{C}_{a_i}\mid_\unrhd& = & \lbrace a_k  \bigcup C_{ik} \mid \forall \; k \; \exists C_{i\unrhd k} \rbrace \\
\mathcal{C}_{a_i}\mid_\rhd& = & \lbrace a_k  \bigcup C_{ik} \mid \forall \; k \; \exists C_{i\rhd k} \rbrace \\
\mathcal{C}_{a_i}\mid_\gg& = & \lbrace a_k  \bigcup C_{ik} \mid \forall \; k \; \exists C_{i\gg k} \rbrace
\end{eqnarray}
By using the legal analytical concept of {\sl privacy spheres}, a functional definition of privacy, confidentiality can be implemented. These are the building block upon which the control of the information emitted by the sender can lead to more complex human concepts like sovereignty and reputation.  
\subsubsection*{Connecting autonomous principals $a_i$}
\label{subSubSec:connections}
The existence of peer-to-peer relationships between autonomous principals is a prerequisite of the model. Pre-digital transformation governances regulated these connections. As explained in the introduction (see page \pageref{sec:introduction}), in post-digital transformation the intrinsic properties of digital communication have to be included in governances to provide security for the ecosystems.  
\par Connections are one-to-one relationships only $1\longrightarrow1$ defined on the information network ${\cal W}$. This ensures that we can bind the connection with the governance regulating the interaction between two entities. 
\par The legal analytical concept of privacy sphere includes at least a public, private, and intimate sphere. To provide a functional definition of privacy spheres that differentiates between these three types, we attach a {\sl confidentiality level} to each message. When the sender agent $a_i$ is autonomous, the distributed governance model assumes $a_i$ can decides on the confidentiality level of a message send. The receiving entity is aware of this confidentiality level either by a direct notification from the sender or from the context in which the message is exchanged. Therefore, the model can handle situations where both parties interact in different context.
\par
Information is composed of data and extrinsic properties that impacts the information exchange events. The three confidentiality level are intrinsic property of the connection under the control of the {\sl sending} autonomous principal. Flagging an information item is, often necessary, but not sufficient. To capture the three types of confidentiality levels, the information system must also flag the connection  {\sl events}. 
\par
\begin{figure}[!h]
\center
    \includegraphics[width=40mm ]
    {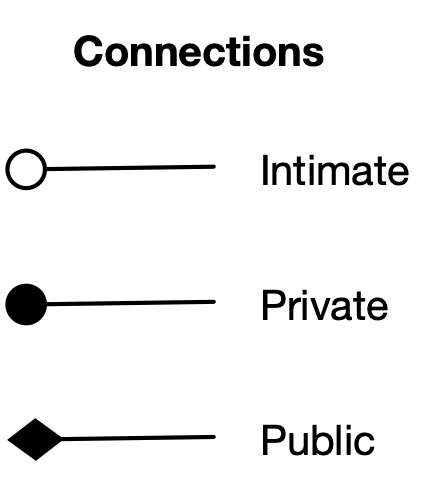}
    \caption{\textit{Core confidentiality levels} 
    \label{fig:confidentialityLevels}}
\end{figure}

Figure \ref{fig:confidentialityLevels} on page \pageref{fig:confidentialityLevels} defines a visual representation of three core confidentiality level.

\par The confidentiality sphere concept imposes the following requirements:
\begin{itemize}
\item[-] {\sl for objects}: flag the confidentiality level at data capture time;
\item[-] {\sl for events}: the sending agent flags the confidentiality level. The receiving entity is aware of the confidentiality level either by a direct notification from the sender our from the context in which the message is exchanged.
\end{itemize}
\subsubsection*{A note on consent management}
\label{subSubSec:consentManagement}
In the physical world, even if the legal concept of {\sl consent} is difficult to precisely define as it relies heavily on the context in which a consent is expected. There are legal protections in place but it is a difficult regulatory domain. When an illicit breach of consent occurs, the correct interpretation of facts provided by the relevant parties relies on the accuracy of the information. In the digital world the secondary usage of data by the receiver has been made easier. Thus managing consent is even harder. Determining a breach of consent faces the same difficulties of the physical world but, due to technological misunderstanding from the users of the information systems, illicit behaviours in the physical space have become standard practice in the digital world.
\par
Managing consent through data usage governed at the level of ecosystems while controlled by the responsible autonomous principal is  one of the main goal of our current research work. A secure digital transformation requires digital consent to evolve from the current agreements based model to a dynamic consent management considering all contextual aspects leading to a digital transaction.
 \par
 A final note on the model's fundamentals. It is important to understand that full digital decentralisation is required for governing fundamental rights of the autonomous principals. Then centralisation through secured platforms can take place where the optimum solution requires it. The distributed governance model is designed to support the exchange of information across platforms. Thus scalability is achieved by transforming the network into a set of interoperable services for a purpose instead of large platforms.

\subsection{Examples\\{\sl Data portability - Causality - Multi-jurisdiction}}
\label{subSec:examples}
To illustrate the application of the distributed governance model, we provide three examples of real-world solutions.
\begin{enumerate}
\item {\bf Data portability \& Asymmetric data usage.} The digital passport use case shows the concept of autonomous principals $a_i$ as the carrier of certified information between ecosystems. This scenario also demonstrates the asymmetric data usage by each ecosystem according to its own needs. 
\item {\bf Causality \& Ordering of events.} {\sl "Data has value when it flows"} is seen through the birth attestation vs. birth certificates. The example deals with the causal order of two events, birth and recognition of birth by the legitimate authority of an ecosystem $E_{\cal{A}}$. This scenario includes experiences collected from a pilot project carried out in east Asia \cite{iRespond:2020,iRespond:2021} and deals with the biometric question.
\item {\bf Multi-jurisdiction \& Data object.} The Covid-19 pandemic has demonstrated the complexity of creating globally accepted proof of vaccination\cite{WHO:2021_07}. This situation demonstrates that today, the digital transformation of the World Health Organisation's vaccination pass is still work in progress .
\end{enumerate}
We provide here elements only addressing the governance aspect of these scenarios.
\subsubsection{Data portability \& Asymmetric data usage \\{\sl Digitalisation of passport}}
\label{subSubSec:digitalPassports}
International travel provides an example of how ecosystems are at the center of a peer-to-peer based distributed governance system.
\par
\begin{figure}[!h]
\center
    \includegraphics[width=80mm ]
    {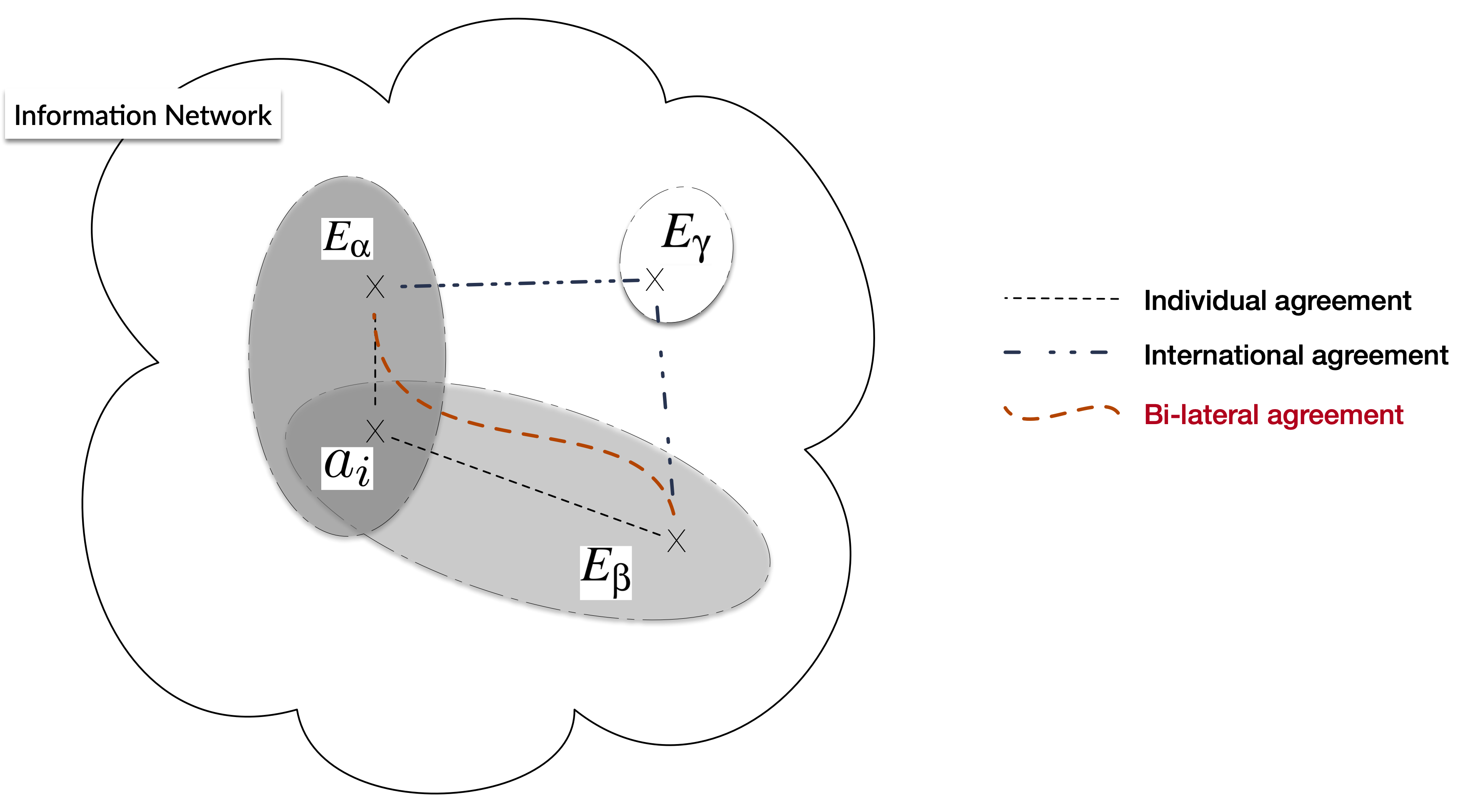}
    \caption{\textit{Distributed governance model for passport usage} 
    \label{fig:passportExample}}
\end{figure}

Using the notation introduced in section \ref{sec:distributedGovernance}, we consider the governance mechanisms for a citizen $a_i$ of country $\alpha$ traveling to country $\beta$. Each country is an ecosystem by itself with a legitimate authority implemented through governmental institutions. $a_i$ as a citizen is a member of the $E_\alpha$ ecosystem and is entitled to receive a passport provided by the relevant governing administration. 
Figure \ref{fig:passportExample} on page \pageref{fig:passportExample} provides a visual representation of information usage asymmetry.
This is a peer-to-peer individual agreement between the citizen and the passport issuing authority. When $a_i$  intends to travel to $\beta$, he enters the $E_\beta$ ecosystem. The recognition of $a_i$ passport by $E_\beta$ will rely on biliteral agreements between the the countries represented by a peer-to-peer agreement between the two ecosystems. In practice, these international agreements are often related to international treaties that both countries adhere to. For example, in the case of international air-travel, this could be for example ICAO represented by the $E_\gamma$ ecosystem in figure \ref{fig:passportExample} on page \pageref{fig:passportExample}.
\par
The model highlights that all interactions are peer-to-peer with the legitimity of the interaction and assurance level required for its authenticity is distributed to each ecosystems' governance. Figure \ref{fig:passportExample} represents the individual and international agreements as straight lines. At the level of passport usage for an individual travel, the individual passport issuance and international agreements are an ambient governance. The agreements are auditably enforced by established rules. The information usage by all parties is symmetrical as its intended usage for a given purpose is transparent.
\par
The mutual agreement between countries to recognise the validity of a passport is more complex. The dynamic context in which the travel of country $\alpha$ citizen $a_i$ to country $\beta$ plays a material role. The model graphically represent this through the usage of a Bézier curve showing the primary self-interest alignment of the parties in a peer-to-peer transaction. In the simple example presented here, it shows that the primary self-interest of country $\alpha$ is aligned with its citizen security while for country $\beta$ it is aligned with protection of the visiting country interests.  Different self-interests lead to different usage of the information being exchanged.
\par
An example where this asymmetry is visible is the absence of global agreements on the travel history that must be recorded in a passport. In pre-digital transformation era, the issuance of a second passport to hide travel in certain countries was a legitimate governed process. The digitalisation of state credentials re-opens the question  and re-inforce the need of governances agnostic to physical or digital implementation.

\subsubsection{Causality \& Ordering of events \\{\sl Birth Attestation vs. Birth certificates}}
\label{subSubSec:birthAttestation}
Causality is a natural concept in the physical world. Events are ordered and a sequence of observations supports a causal logical construction that can be used to determine the liability of an entity in a given governance framework. In the digital world, causality must be reconstructed. The part two and three of this series will provide the technological construction of causal events by first introducing the {\sl unicity} of objects and data model based on directed acyclic graphs \cite{wiki_DAG}.
\par
This reports stays on the governance level and illustrate the importance of event ordering in a case that also illustrate the importance of human-centric governance in digital systems. 
\par
 A birth certificate is a simple but lifelong document that most of us take for granted. However, for many, it is not. {\sl To provide for all a legal identity, including birth registration}, is the United Nations Sustainable Development Goal number 16.9 (see page 25 of reference \cite{UN_SDG:2015}). Stateless newborns count in millions. For these, proof-of-existence through a birth attestation is the first improvement needed. Then a birth certificate can subsequently be issued. Therefore this provides an example of the necessity of event ordering and the importance and risks of digital transformation in the humanitarian sector in the introduction, \ref{subsec:humanitarian} on page \pageref{subsec:humanitarian}.
\par
\begin{figure}[!h]
\center
    \includegraphics[width=80mm ]
    {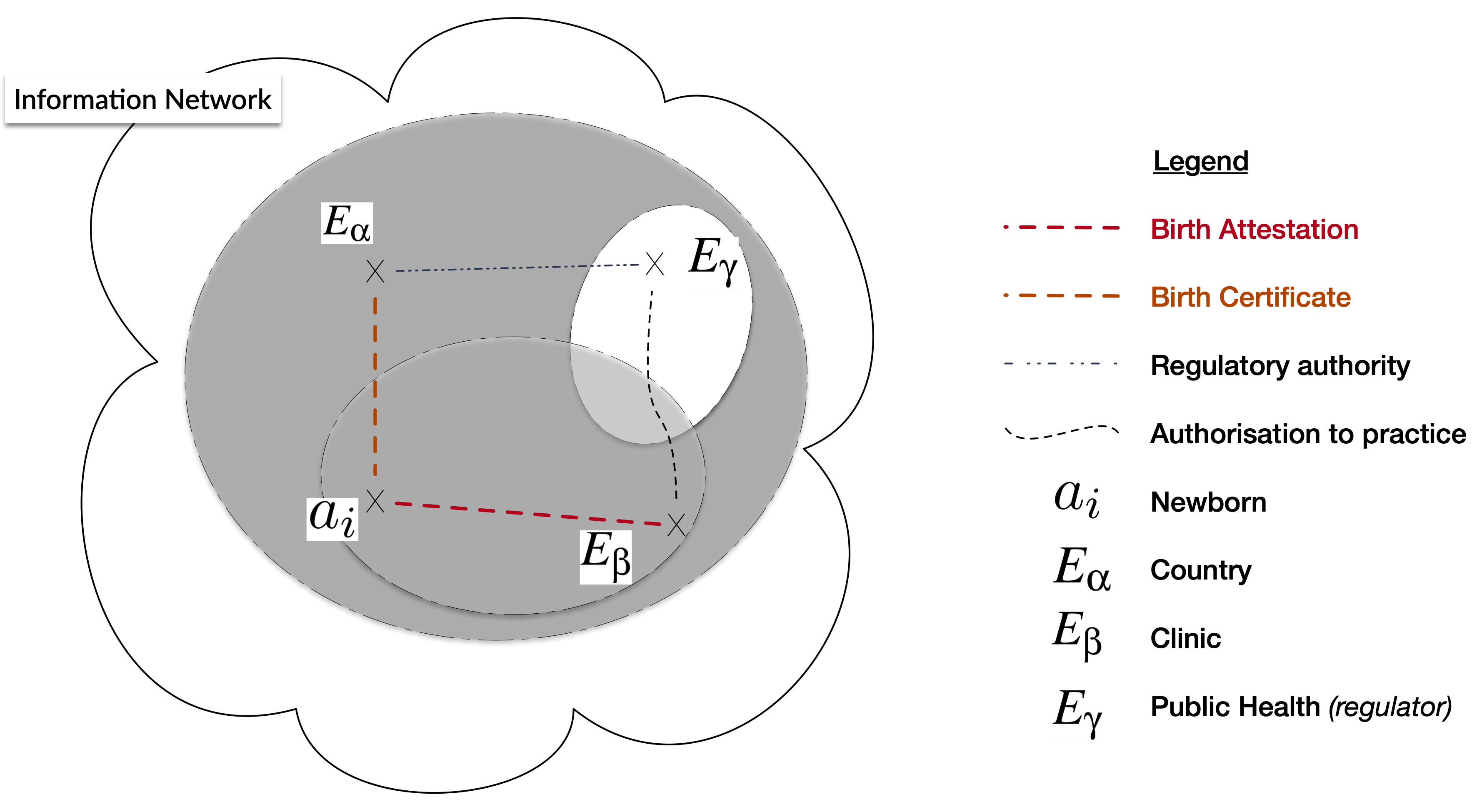}
    \caption{\textit{From proof of existence to birth certificate } 
    \label{fig:BA}}
\end{figure}

In a stable environment, issuing a birth attestation and a birth certificate are closely related processes. The clinic issues the parents certified documents that allow the state to issue a birth certificate. When this occurs in a well-governed environment, as depicted in figure \ref{fig:BA}, the information flow is straightforward, but its digitalisation is not, as it requires system interoperability. Therefore, the benefits of a user-centric governance model are present but not critical. 
On the contrary, the distributed governance model is necessary for an unstable environment to provide a solution protecting the user. Ungoverned digital transformation, even if privacy-protecting, can bring more harm, as recognised in \cite{Cheesman:2020}.
\par
The figure displays the target governance model designed for a pilot project. The autonomous principal is the newborn's mother, and biometric identification is the only persistent root of trust. The humanitarian clinic where the birth takes place issues an attestation of the birth and time. The birth certificate will be issued much later. Multiple intermediaries carry the information and increase the potential misuse of the information (voluntary or not). Digitalisation increases these risks compared to a paper-based solution where the information flow is easier to contain. 
\par
From a governance perspective, this example illustrates why the autonomous principal must remain the root of trust when information moves from one ecosystem to another. In digital terms, this implies that the underlying technologies related to biometry play a vital role.  

\begin{figure}[!h]
\center
    \includegraphics[width=80mm ]
    {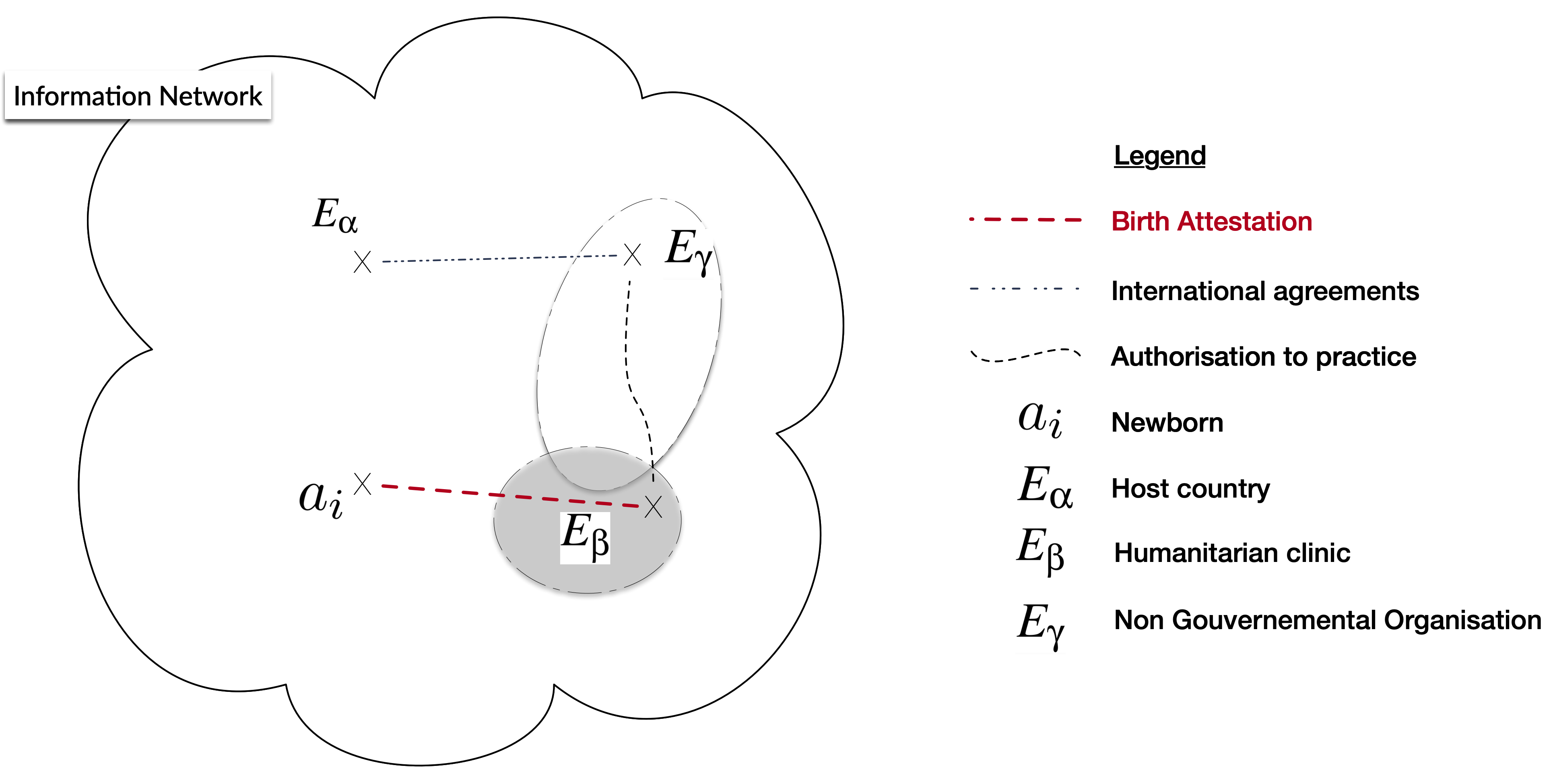}
    \caption{\textit{Stateless proof of existence} 
    \label{fig:BAHumanitarian}}
\end{figure}

Figure \ref{fig:BAHumanitarian} on page \pageref{fig:BAHumanitarian} shows the difference in governance for humanitarian action from the perspective of the vulnerable individual that can not rely on, or needs to be protected from a state ecosystem.
\subsubsection{Multi-jurisdictions\\{\sl Dealing with multiple governance }}
\label{subSubSec:multiJurisdiction}
The prevailing platform-centric paradigms have resulted in an oversimplified approach to governance models, failing to address current demands adequately. As elucidated in the introductory section, our society evolves within an intricate web of rules and regulations to safeguard human and community values. With an escalating volume of digital information exchange, each transaction must receive protection beyond data protection regulations. Privacy laws preserve fundamental freedoms in general. More contextually relevant regulations must be employed to ensure safety. The case of healthcare and the quandary of categorising software as a medical device underscores that contemporary digital systems must adhere to a plethora of regulations surpassing those dictated solely by the platforms hosting the data. Therefore, within the digital realm, a platform-driven data governance framework aligned with local data protection statutes must be revised; multiple governance mechanisms' coexistence is imperative.
\par
We elucidate the approach of the distributed governance model in managing multifarious governance structures, exemplified through two archetypal scenarios: i) Operations of a private multinational corporation spanning diverse nations and ii) Activities of a supranational organisation operating on an international scale. We formulate the general concepts below and leave specific implementation to further publications.
\par
\begin{figure}[!h]
\center
    \includegraphics[width=80mm ]
    {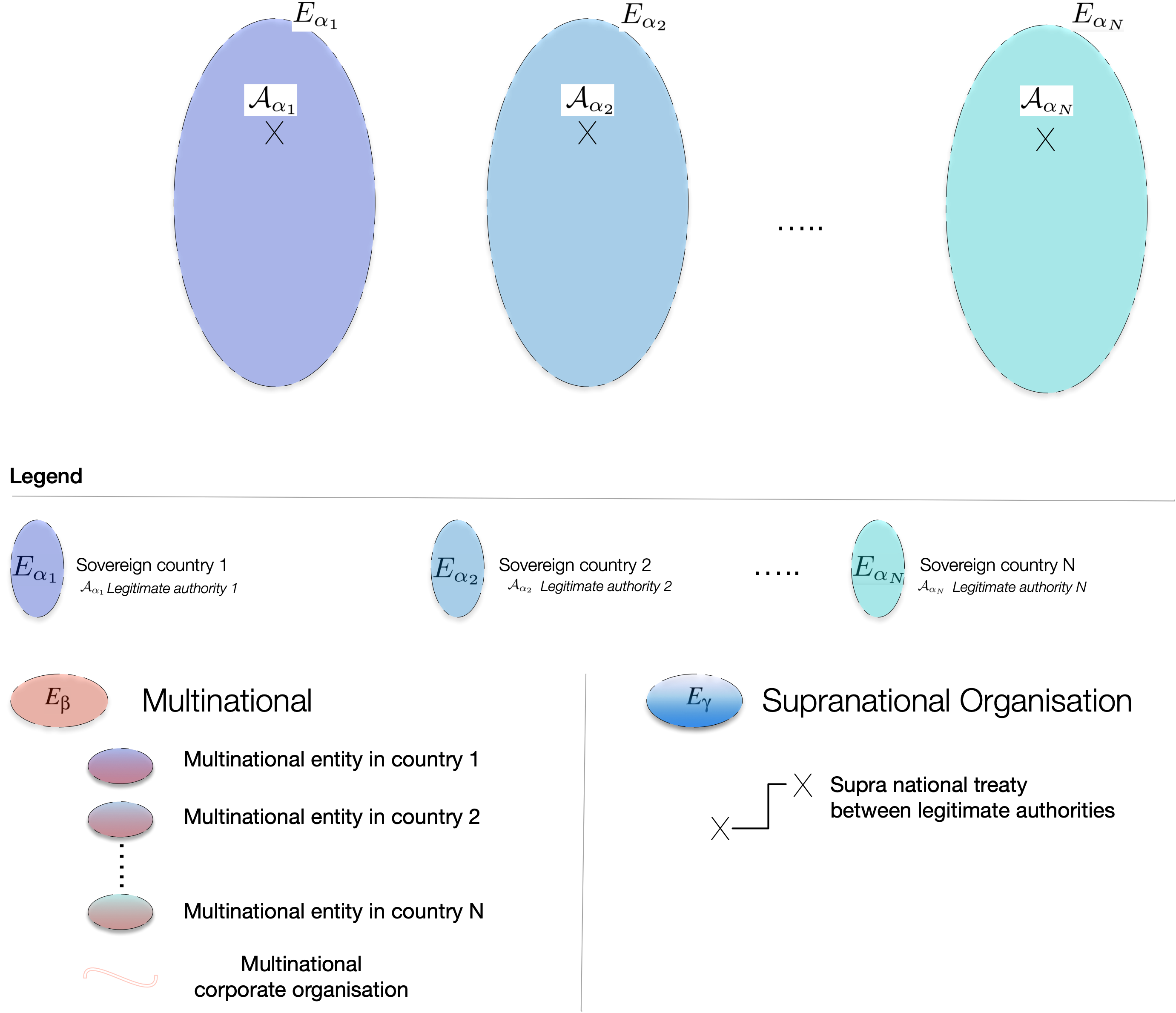}
    \caption{\textit{Multi-sovereign entities set-up} 
    \label{fig:legendMultGovernance}}
\end{figure}
\begin{figure}[!h]
\center
    \includegraphics[width=80mm ]
    {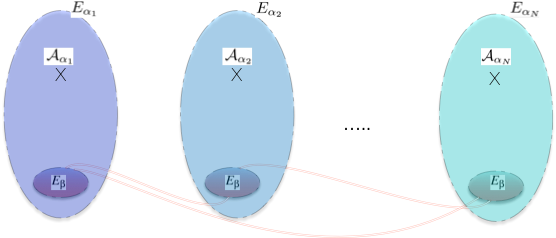}
    \caption{\textit{Multinationals operating in $N$ countries} 
    \label{fig:multGovernanceMultinationals}}
\end{figure}

Consider a set of $N$ sovereign countries, each represented by an independent ecosystem denoted as $E_{\alpha_i}$, alongside a legitimate authority $\cal{A}_{\alpha_i}$. Each country operates within its distinctive set of rules and is free to form agreements, with other sovereign entities. This conceptualisation is depicted schematically in \ref{fig:legendMultGovernance} on page \pageref{fig:legendMultGovernance}.
 A multinational corporation is represented by its self-contained ecosystem $E_\beta$. The corporation assumes the role of an autonomous principal, denoted as $o_\beta$, potentially represented by a legitimate authority such as a board of directors. In this scenario, $o_\beta = \cal{A}_\beta$ holds true. The corporation's purview is whether to engage in operations within a particular country. Within a country, the corporation's presence is reflected through an ecosystem $E_{\beta_i}$, operating within the confines of country-specific rules $E_{\alpha_i}$, and interlinked with other corporate entities (including the headquarters) through agreements.  \ref{fig:multGovernanceMultinationals}  (refer to page \pageref{fig:multGovernanceMultinationals}) provides a schematic description. Essentially, each $E_{\beta_i}$ is subject to the jurisdiction of country $E_{\alpha_i}$ while being controlled, at least partially, by the corporation $E_\beta$.
\par
From a governance perspective, this portrayal of a multinational corporation as an assemblage of distinct ecosystems interconnected through internal agreements underscores the necessity to adhere to the laws and regulations of the respective sovereign countries where operations the corporation conducts its business. Consequently, the permissible activities of the corporation may substantially vary across different countries.
In the case of a supranational entity (e.g., ICAO, United Nations, ICRC), the governance landscape differs markedly. Each sovereign country decides whether to join or abide by the governance framework of the supranational entity. A treaty or charter to which member states adhere embodies the legitimate authority of a supranational entity. These treaties often mandate equal treatment for all members, binding the supranational entity to provide equitable treatment to each sovereign country. Employing the configuration defined in figure \ref{fig:multGovernanceMultinationals} (refer to page \pageref{fig:multGovernanceMultinationals}), an ecosystem operating beyond the jurisdiction of sovereign countries conceptualises a supranational organisation $E_\gamma$. Each $E_{\alpha_i}$ is presumed to be a member of the supranational entity $E_\gamma$. The external location of the legitimate authority of $E_\gamma$ relative to the sovereign countries $E_{\alpha_i}$ enables the establishment of local immunities and other privileges congruent with the mission of the supranational entity. Another instance, as observed in the context of ICAO and the international use of passports, is explored in section \ref{subSubSec:digitalPassports} on page \pageref{subSubSec:digitalPassports}. Here, the supranational entity's mission involves ensuring passports interoperability across its member states. 
\begin{figure}[!h]
\center
    \includegraphics[width=80mm ]
    {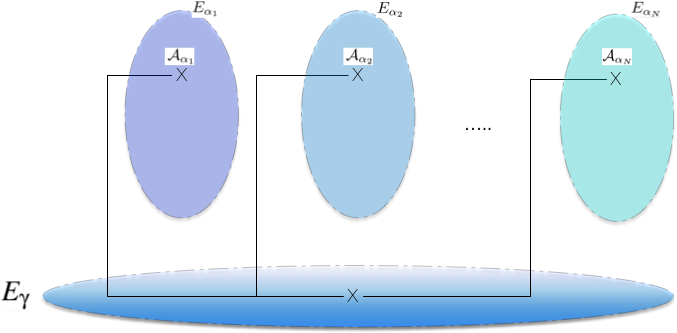}
    \caption{\textit{Supranational entities} 
    \label{fig:multGovernanceSupranationals}}
\end{figure}
The distributed governance model facilitates precise delineation of the interrelationships between sovereign structures. The model's objects also enable the practical implementation of multiple governance frameworks in digital information systems functioning across diverse jurisdictions.
\par
Consider the Council of Europe's Convention 108+ and its dynamic management of the list of signatory states as an illustrative example. This list encompasses the sovereign entities that have ratified the treaties and incorporate vital details essential for treaty implementation. Beyond signature, ratification, and entry into force dates, this list accommodates country-specific annotations concerning reservations, declarations, derogations, denunciations, authorities, territorial applications, objections, and communications. In a distributed governance network, every node representing an autonomous principal subject to the governance of Convention 108+ can propagate the requisite governance particulars pertinent to a given transaction. Similarly, a recipient node can solicit the governance status of a counterparty and enable a verification mechanism commensurate with the required level of assurance for a specific data usage assurance. Moreover, within this framework, the administrative authority (e.g., the Council of Europe) can disseminate modifications through machine-readable notifications accessible to all participating nodes without necessitating an intermediary administrative platform.

%% file: table-cv2-notationAbstractGov.tex
\begin{table}[tb]
\centering
\caption{Notation: Abstract Governance}
\label{table:notationAbstractGov}
\begin{tabular*}{\columnwidth}{@{}l@{\hskip4pt}l@{}}
	\toprule
	$a_i$ & a generic  autonomous principal\\ 
		& $h_i$ an individual human \\ 
		& $o_i$ an organisation \\ 
		& $g_i$ a governing entity\\ 
		& $i$ unique lowercase latin indices for autonomous \\
		& principals. Indice $i$ is \underline{independent} of ecosystem \\
	$E_{\cal{A}}$ & an ecosystem\\
	& $E_{\cal{A}} = \left( a_\alpha, \cal{A}, \cal{A}^* \right)$ \\
	$\alpha$ & lowercase greek indices to label ecosystems\\
	& $\alpha = 1,\ldots, N_{\rm E}$\\
		$N_{\rm E}$ & total number of ecosystem potentially interacting in $\cal W$ \\

	$a_\alpha$ & population vector of ecosystem $E_\alpha$\\
	&$a_\alpha=\left( a_1,\ldots,a_{{N_{{\rm E}_\alpha}}} \right)$\\
	$A_\alpha$ & legitimate authority of ecosystem $E_\alpha$\\
	$Ad_\alpha$ & administrative framework of  $A_\alpha$\\
	$N_{{\rm E}_\alpha} $ & total number of autonomous agent in $E_\alpha$\\
	$\sigma_\alpha$ & a generic non-autonomous agent (e.g. a thing) \\
	& $\alpha$ lowercase greek indices for non-autonomous entities\\
	$\cal W$ & the world\\ 
	\bottomrule
\end{tabular*}
\end{table}

%% file: table_networkCharacteristic.tex
\begin{table}[tb]
\centering
\caption{Network Model Notation}
\label{table:networkModel}
\begin{tabular*}{\columnwidth}{@{}l@{\hskip4pt}l@{}}
	\toprule
	$\cal C $ & characteristics ${\cal C} = \left\{ {\cal C}_A,{\cal C}_D\right\} $\\ 
			& $A={\rm Authentic}$ \\
			& $D={\rm Deterministic}$ \\
	${\cal O}_T$ & transient object ${\cal O}_T = \left\{ {\cal O}_{Tc},{\cal O}_{Tf}\right\} $\\ 
			& $c={\rm Credential}$ \\
			& $f={\rm Form}$ \\
	${\cal O}_P$ & persisten object ${\cal O}_P = \left\{ {\cal O}_{Pr},{\cal O}_{Ps}\right\} $ \\
			& $r={\rm Record}$ \\
			& $s={\rm Schema}$ \\	 
	$\cal E$ & element ${\cal E} = \left\{ {\cal E}_c,{\cal E}_a \right\} $ \\
			& $c={\rm claim}$ \\
			& $a={\rm attribute}$ \\	 
	$\cal S$ & state ${\cal S} = \left\{ {\cal S}_\alpha,{\cal P}_{\bar{\alpha}} \right\} $ \\
			& $\alpha={\rm active}$ \\
			& $\bar{\alpha}={\rm passive}$ \\	
	$\cal P$ & process ${\cal P} = \left\{ {\cal P}_\beta,{\cal P}_{\bar{\beta}} \right\} $\\
			& $\beta={\rm entry}$ \\
			& $\bar{\beta}={\rm capture}$ \\ 
	$\cal R$ & Context ${\cal R} = \left\{ {\cal R}_\rho,{\cal R}_{\bar{\rho}} \right\} $\\
			& $\rho={\rm factual}$ \\
			& $\bar{\rho}={\rm objectual}$ \\ 	

	\bottomrule
\end{tabular*}
\end{table}

%% file: table_exampleMessage.tex
\begin{table}[tb]
\label{table:exampleMessage}
\caption{Example of messages}
\begin{center}
\begin{tabular}{rl}
	\toprule
	$ h_i {> \atop \longrightarrow} h_j $ & an individual speaking to another \\
 	&  \\
	$ h_i {>< \atop \longleftrightarrow} h_j $ & two individuals present  \\
 	& in the same room \\
  	&  \\
	$ h_i {> \atop \longrightarrow}  Ad_\alpha$ & an individual sending \\
 	& a request to an administration\\
	\bottomrule
\end{tabular}
\end{center}		
\end{table}

%% file: CV3-ConsensualVeracity.tex
%
\section{The Dynamic Data Economy alternative }
\label{sec:consensualVeracity}
\subsection*{Introduction}
\label{CV3-subSecIntro}
The distributed governance model of section  \ref{sec:distributedGovernance}  provides a framework for the functional design for the governance of ecosystems and their interactions with other ecosystems. The model relies on interacting agents (autonomous principals and ecosystems) and applies to digital and non-digital interactions. As a result, the model can approach digital governance problems starting with rules and regulations existing prior to digital transformation.
\par
This section presents an alternative framework for digital interactions coined by Dynamic Data Economy (DDE). DDE is a data-centric framework that relies on four data domains: Objects, Events, Concepts and Actions \cite{Knowles202305}. 
Decentralised software technologies underpin the DDE framework to protect the data objects' integrity and the data events' authenticity. Collectively, they provide the data infrastructure upon which the distributed governance model presented here digitally enables human concepts and actions.
\par
Specifically, the {\sl concept domain} regroups the knowledge elements upon which a given ecosystem builds its governance to access data. For example, this is where the regulatory instruments are required for an ecosystem's administration. The action domain represents the economy of an ecosystem where the interaction of autonomous principals (i.e. economic actors) generates value. For example, this is the domain where algorithms live (e.g. Artificial Intelligence)\footnote{an application to the distributed governance model to the regulation of AI in a DDE framework is the subject of an upcoming publication}
\par
Within the DDE framework, the distributed governance model allows a coherent integration of existing principles, rules, and regulations governing a community (i.e. ecosystem) into a digital representation. As a result, the model extends the concept of data governance \cite{Ladley_2012} driven by data protection regulations now available in over 130 countries \cite{UNCTAD_2023} to any concepts that a community adheres to. Therefore the model helps design information systems operating across multiple jurisdictions. The autonomous principals are the liable entities navigating across ecosystems.
\begin{figure}[h!]
\center
    \includegraphics[width=90mm ]
    {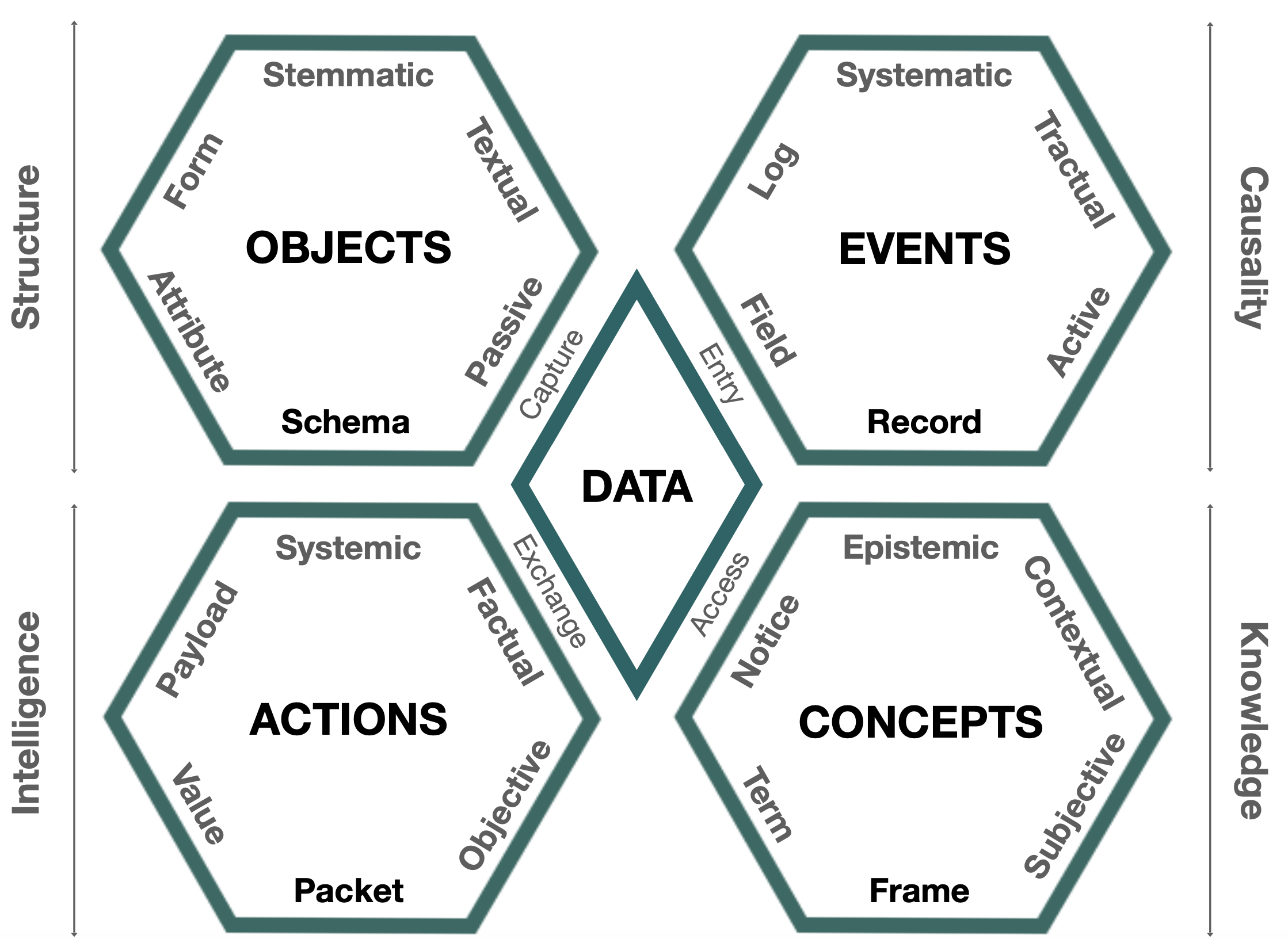}
    \caption{\textit{Four domains of a Dynamic Data Economy \cite{Knowles202305}} 
    \label{fig:DDEdomains}}
\end{figure}
\par
From a technological perspective, multi-jurisdiction governance for data management requires semantic interoperability and factual authenticity of events across different ecosystems. First, in subsection \ref{subsec:objectualIntegrity}, the Object domain leads to the concept of Objectual Integrity used to define deterministic objects upon which different actors can agree on the definition. Then, subsection \ref{subsec:factualAuthenticity}  introduces Factual Authenticity as the basis for a digital equivalent of causality emerging from the Event domain. As a result, DDE provides an instance of figure \ref{fig:informationNetwork} on page \pageref{fig:informationNetwork} and an explicit construction of digital space where technological assurance blends into a Human context where trust can emerge.
\par
Both domains are intimately entangled to produce a digital infrastructure that secures the intrinsic properties of data independently of its location. Section \ref{subsec:consensualVeracity} builds on the data-centric infrastructure to develop a sense of consensual veracity with the outputs of information systems. The section ends relates data-centric infrastructure to the Human decision-making process in a digital environment. This section concludes with the primary objective of this work, the agency question in the digital age.
\subsection{Objectual Integrity\\{\sl a necessity for structured data}} 
\label{subsec:objectualIntegrity}
In a peer-to-peer information system, the integrity of a data object captured by an information system must rely on mechanisms under the control of the receiving party. These mechanisms must ensure that the object is temper resistant and morphologically structured. They must ensure that the data and its definition are securely communicated for secure communication. 
\par
A foundational protocol is necessary to morphologically structure data so that actors can unanimously agree on its meaning. Consequently, the challenge of dealing with semantics arises, posing a complex issue in computer science. In a significant article titled "The Semantic Web", published in May 2001 in Scientific American, Tim Berners-Lee, James Hendler, and Ora Lassila pioneered a web of data that could be shared and processed by both machines and humans. Over time, semantics has become a vibrant area of research, encompassing concepts such as RDF (Resource Description Framework), XML (eXtensible Markup Language), linked data, and OWL (Web Ontology Language) \cite{Antoniou,Heath,Allemang,WuHuajunCudre_2012}. Additionally, it emphasises the importance of creating shared vocabularies and ontologies that facilitate machines in comprehending and interpreting data in a standardised manner. 
\par
These advances provide excellent tools to organise and search data at the scale of today's Internet. Nevertheless, these mechanisms are not designed to capture the context in which data are produced and captured. Today's digital transformation of society as a whole requires that the structure of data is preserved across platforms and multiple jurisdictions.
\par
Graph technologies focus on analytics and querying (property graphs) and data integration (RDF) but often disregard the data harmonization process, leaving no guarantee that the structural and contextual integrity of the data is maintained. However, controlling the data transformation process requires structuring the data. It starts with data capture by purpose-based service providers (services that capture data for a specific purpose). Then insights-based service providers (those seeking existing data for analytics and insights) can process and prepare datasets for machine learning to maintain objectual integrity through its evolution.
\par
In a distributed governance network, the requirements for Objectual Integrity that support peer-to-peer interactions are:
\begin{description}
\item {\bf Rich contextual metadata:} The captured context and meaning (the "metadata") for all payloads MUST be rich enough to ensure complete comprehension by all interacting actors, regardless of written language.
\item {\bf Structured data form:} Data governance administrations $Ad_i$ MUST publish structured data capture forms, specifications, and standards, driven by member consensus for a common purpose or goal that will ultimately benefit the global citizens and legal entities they serve.
\item {\bf Harmonised data payload:} There are two areas of distinction to consider. {\sl Data harmonisation} involves transforming datasets to fit together in a common structure. {\sl Semantic harmonisation} ensures that the meaning and context of data remain uniformly understood by all interacting actors, regardless of how it was collected initially. Harmonised payloads are a MUST for multi-source observational data to ensure that the data is used for machine learning and Artificial Intelligence.
\item {\bf Deterministic Object Identifiers:} Object identifiers MUST associate with a cryptographic hash of digital content. A hash value is a deterministic fingerprint for digital content. If any operation's result and final state depend solely on the initial state and the operation's arguments, the object is deterministic. In other words, any resolvable object via its digest is deterministic.
\end{description}

{\sl Decentralised Semantic}\cite{knowles2022} is the term coined to describe an architecture meeting these requirements. It takes the form of a standard layered schema architecture for segregating task-oriented objects. Decentralised semantics is ontology-agnostic, offering a harmonisation solution between data models and data representation formats. It provides a secure digital object definition upon which a distributed governance is applied and the contextual elements to resolve privacy-compliant data sharing between servers, networks, and across sectoral or jurisdictional boundaries. Internationalisation and the dynamic presentation of transient objects are examples where decentralised semantic use.
\par
One of the core characteristics and advantages of decentralised semantics is that different actors from different institutions, departments, sectors or jurisdictions can acquire control of specific task-oriented objects within the same semantic structure by binding an authorisation credential to that object. 
\begin{center}
\begin{figure}[h]
    \includegraphics[width=\linewidth,
   ] {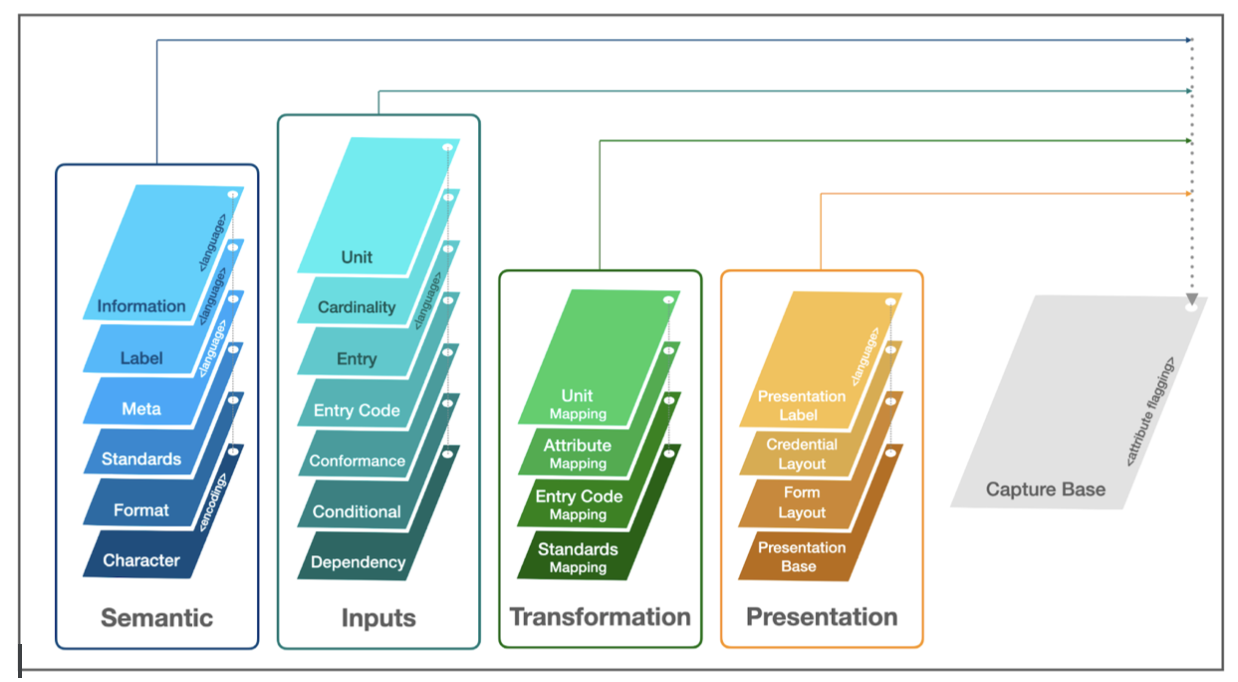}
    \caption{Layered Architecture for task-oriented objects}
    \label{fig:layeredArchitecture}
\end{figure}
\end{center}
An implementation of a layered architecture meeting the requirements of decentralised semantics is the Overlays Capture \cite{Knowles2022b} being OpenSourced by the Human Colossus Foundation and accessible through this reference \cite{HCF_OCA:2022}. 
\subsection{Factual Authenticity\\{\sl Towards deterministic digital causality}} 
\label{subsec:factualAuthenticity}
Suppose one can verify the provenance of data and its integrity. In that case, one can assume that data is authentic and can be used in decision-making. The previous section dealt with data integrity; this one focuses on data provenance. In computer science, data provenance links to the process of authentication. However, authentication of the issuer is necessary but not sufficient, and more is needed. Unfortunately, we often forget that the chain of events, causality, has to be built in a digital system. Causality is essential to determine's an actor's liability in the human world that. The same mechanisms must be reconstructed in the digital space. Therefore, the usual time stamping mechanisms need to be more secure to ensure a verifier (i.e. a judge) of the integrity of a chain of events using specific data structures like acyclic graphs (DAG)\cite{Pearl:2009}, for example. 
\par
This section briefly introduces the concepts of immutably ordered events and recently developed decentralised key management architectures meeting the requirements of data provenance for distributed governance.
\par
In the digital space, everything starts with {\sl Data Inputs}. These sequences of bytes provide information to a computer at a point in time. In computer science, a factual event is an action or occurrence identified by a program that has significance for system hardware or software. Stored sequences of bytes are what we commonly call "data".
\par
Data input distinguishes itself from {\sl Data Entry}, the process of transcribing information into an electronic medium such as a computer or other electronic device. The process entails storing state changes as recorded events to determine the authenticity of the data origin (the "source"), its status, and where it moves over time. It is essential to notice that the ordering of the event must be immutable and verifiable. In a balanced digital network, data entry requires append-only logs to accompany signed data inputs to identify the origin and creation of authentic events at recorded moments so that the inputted data can be considered factual.
\par
So, "Authenticity" relates to unique events. In the digital space, unicity emerges from cryptographic unicity. These considerations lead to the formulation of two principles followed by a Dynamic Data Economy implementation:
\begin{description}
\item {\bf Authentic data events:} Public/private key pairs provide the underpinning for all digital signatures, a mathematical scheme for certifying that event log entries are authentic. All recorded events MUST be associated with at least one public/private key pair to be considered authentic. 
\item {\bf Verifiable event identifiers:} Data provenance provides a historical record (an "event log") of the data and its origins. All event identifiers MUST be cryptographically verifiable to ensure data provenance, which is necessary for addressing validation and debugging challenges.
\end{description}

With a mechanism providing authentic and verifiable events, the governance question translates into the problem of ensuring a strong link between the autonomous principal $a_i$ and its dual digital representative $a_i^\star$. Therefore, the technical problem to solve is the control of $a_i^\star$'s identifiers on a data network. Once addressed, the strong binding between $a_i^\star$ and a human ($a_i$) can be tackled. This will be the subject of subsequent publication as the matter deals with biometrics and other forms of authentication with physical devices.
\par
This control of digital identifiers is a thorny technology question underlying global and local initiatives related to so-called {{\sl digital identities}. Fueled by privacy, governance and economic interests, many avenues are pursued. For example, the Self-Sovereign Identity (SSI) \cite{ReedPreukshat2020, YoungVescent:2018}. movement, by focusing on decentralised authentication, is gaining traction across many domains, including for governmental use \cite{swiss:2021}.  
\par
However, it is only with the emergence of the Key Event Receipt Infrastructure (KERI) \cite{Smith:2020_10,Smith:2019v2} in the early 2020s that a genuinely decentralised identifier system (DKMS) scalable to real-world problems became available. In a nutshell, KERI enhanced the security of identifiers on two dimensions.
\par
First, by maintaining a strong cryptographic binding between an identifier and its controller without an administrative intermediary, KERI removes a weakness of the current cryptographic key management systems (KMS) underpinning authentication protocols. Therefore, with KERI, security for the system concentrates on securing the private keys of the controller. To this extent, KERI introduces key logs and a key pre-rotation mechanism where the next keys are readily available to secure key rotation events.
\par
Second, KERI avoids the requirement of  a centralised system (e.g. registries, DLTs) for the promulgation of public keys by letting a set of network agents chosen at the discretion of the controller or verifier. The controller of the identifier determines the visibility of his public keys log through a set of agents. On the other hand, the verifier determines the adequate set of agents in function of the assurance level required for signature verification. The system also provides an additional set of agents to address the internal risk of a fraudulent controller (i.e. duplicity risk). Therefore, the controller and verifier of an identifier can separately define the identifier's security according to their respective need or level of assurance. KERI establishes a truly decentralised authentication system. 
\par
KERI is developed as an open-source standard at the Linux Trust over IP Foundation (ACDC), Decentralised Identity Foundation (DIF) and Human Colossus Foundation (HCF), where Python and Rust reference implementations are produced. In addition HCF initiated the development of authentication libraries integrating KERI and other authentication protocols to establish interoperability between platforms. The EU Horizon 2020 project DKMS-4-SSI initiated this work.
\par
At the time of writing, in 2022, the first implementation of KERI in a production environment was done by the GLEIF foundation for the issuance of verifiable legal entity identifiers.
\subsection{Decision-making in the digital age\\{\sl Consensual Veracity of information sources}} 
\label{subsec:consensualVeracity}
\par
\begin{center}
\begin{figure}[h]
    \includegraphics[width=\linewidth,
   ] {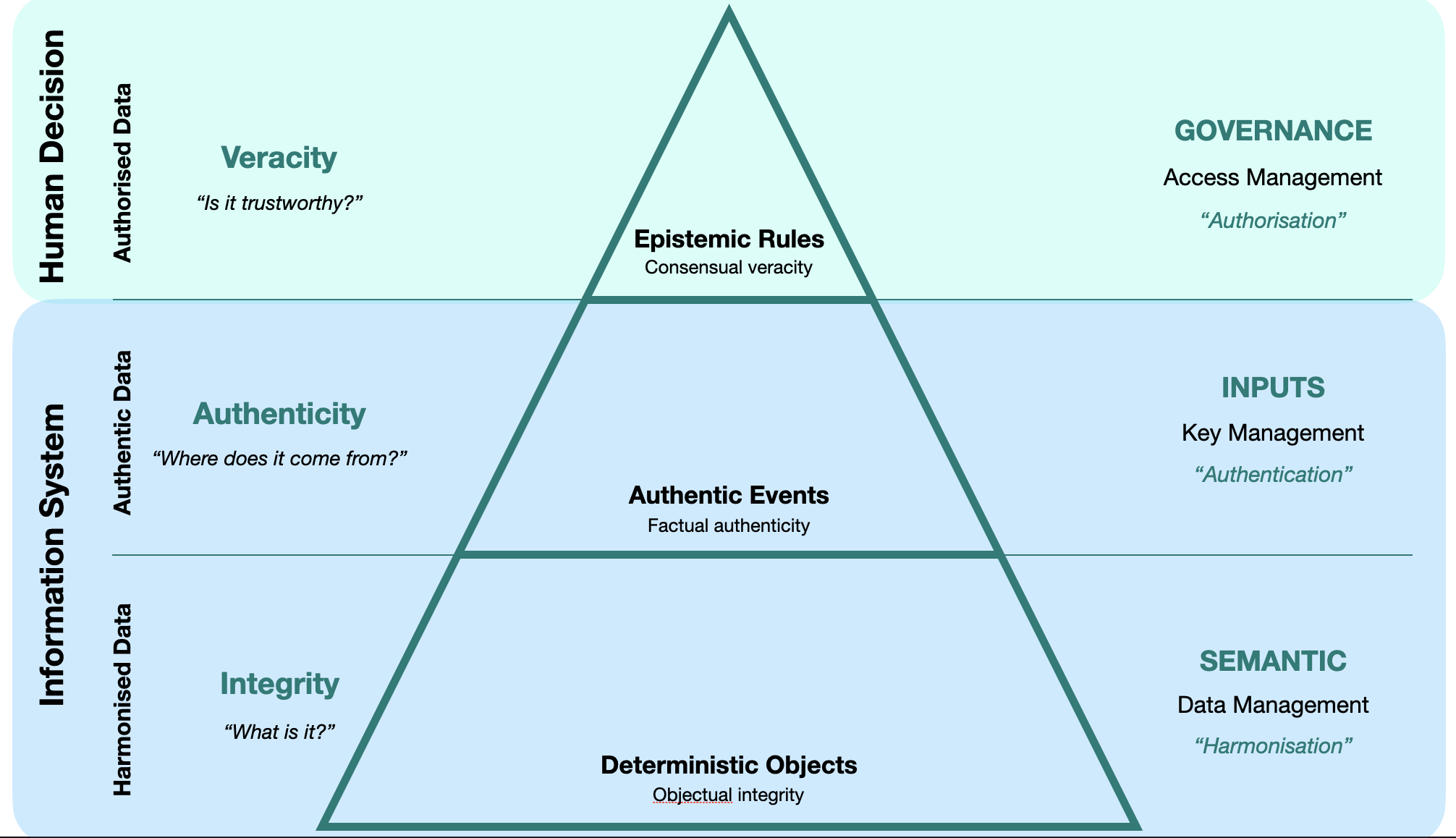}
    \caption{Human Decision Making resting on decentralised architecture}
    \label{fig:}
\end{figure}
\end{center}
We coined the term {\sl Consensual Veracity} to underline that the assessment of veracity results from a consensus the decision-maker must reach with its different information sources providing the rationale basis of a decision. The ultimate goal of an information system is to help individuals, organisations, and governing bodies make better decisions based on accurate information. Deciding means dealing with uncertainties and thus differs from scientific understanding. Identifying the relevant features pertinent to a decision is crucially dependent on the context of the decision. AI and advanced analytics undoubtedly bring a more refined understanding of a situation by gathering more data. Nevertheless, more than accessing large amounts of data to improve decision-making is required. The context in which the decision is made and the context in which the data is captured is often more important than the data itself.
\par
The distributed governance model recognises that a decision-maker will rely on multiple information sources, some digital and some not. Thus, to assess the veracity of the relevant features, the decision-maker must build a personal consensus from information originating from different sources and ecosystems. 
\par
From the most straightforward device equipped with advanced analytics to sophisticated AI, today's information systems are driven by scientific accuracy through algorithmic designs that can be traced back to the work of Fisher  \cite{Fisher:1956, Fisher:1973} on scientific inference. However, dealing with uncertainties is still an active domain of research. In information systems, preserving context requires merging meaning from different sources, which is the purpose of the emerging concept of decentralised semantics. Semantics also contributes to data accuracy when traceable contextual elements can be preserved in the data transformation process. For example, inferential models\cite{Martin:2015} that attempt to extract relevant properties from data-dependent probabilities will benefit from context-preserving technology when isolating relevant features.
\par
Within a distributed governance framework, objectual integrity \ref{subsec:objectualIntegrity} and factual authenticity \ref{subsec:factualAuthenticity} provide transparency on data provenance. The fractal property of the model allows ecosystems to be structured and interlaced. Thus data accuracy can be done {\sl within and across ecosystems}. The decision-maker shapes the contextual elements needed.
\par
Within an ecosystem $E_\alpha$, data veracity is how accurate or truthful the data is, including data quality and how trustworthy the data source, type, and processing are. The ecosystem's governance authority fixes the rules enforced by its administration, often in terms of levels of assurance, for example. This leads to the following requirements for the data governance administration $Ad_\alpha$.
\begin{description}
\item {\bf Reputable data actors:} Data governance administrations MUST exercise vigilance to ensure that all ecosystem participants involved in digital interactions under their administrative control are reliable and trustworthy.
\item {\bf Accountable data governance:} Data governance administrations MUST assume responsibility for the veracity of epistemic rules for safe and secure data sharing for the ecosystem participants and legal entities they serve. 
\item {\bf Searchable distributed databases:} Data governance administrations MUST house at least one distributed database that insights-based service providers can utilise for structured criteria searches and data requests.
\item {\bf Monitored data requests:} Data governance administrations MUST ensure that domain experts constantly monitor any dynamic search engine targets under their administrative control to protect members against unethical or sensitive incoming data requests.
\item {\bf Consensual policy:} Privacy rights, data governance policy, and licences MUST provide the legal basis for safe and secure data sharing within and between sectoral or jurisdictional ecosystems for a particular purpose.%
\end{description}
\subsection{Agency in digital space\\{\sl For a purpose based digital economy}}
\label{subsec:agencyProblem}
Trust is a human concept that is at the core of the Principal - Agent problem. Through trust the principal lets another party take control of transactions that should bring a benefit to the principal. The same logic applies to information systems where an autonomous principal $a_i$ gives information to a system for a purpose and expects the agent to fulfil the purpose, nothing more. {\sl Trust} does not apply to machine, {\sl assurance} should be used. But contrary to machines, digital information systems deal have become intrusive and also provider of intimate information used for decision making and these decision can not be surveilled or influenced by external parties.
\par
DDE provides an instance of the distributed governance model that rests on a technological digital architecture designed to preserve privacy and data accuracy for better decision making. Within DDE, the autonomous principal $a_i$ interacts with the information system through a Self-Actioning System (SAS) and an array of use case specific user interfaces. Collectively, they embody $a_i^\star$ the digital self, dual to $a_i$. A strong binding between $a_i$ and $a_i^\star$ provides the anchor upon which the digital liabilities can be assigned.
\subsubsection*{Digital-Self as the missing element} 
Standards and governances already exist for today's non-digital communication channels. Therefore, we should focus on defining the functional implementation confidentiality level for digital information systems to build digital governance.

When participating in an information system, an entity should be able to specify the confidentiality level when exchanging or disseminating information. Regardless of the connection channel, the entity must retain control over the confidentiality level. To meet this implementation requirement, we need a fully peer-to-peer (i.e., decentralized) authentication of the identifiers managed by the entity and fully decentralized semantics to ensure that flagging the confidentiality level does not compromise the integrity of the exchanged digital objects. The architectural solutions presented in this section meet these requirements and are actively developed. The KERI-based protocols enable decentralized authentication, and layered schema architectures like OCA offer decentralized semantics solutions. 

These architectures enable the functional definition of digital $a_i^\star$, a {\sl digital-self}, representing the autonomous principal  $a_i$ for individuals, organizations, or governing institutions. Leveraging decentralized authentication and semantics establishes a strong binding between $a_i$ and $a_i^\star$. This binding is the link between the information and data networks. As a result, a digital representation of the autonomous principal is a digital self. 
Through the binding, we have a dual representation of the autonomous principal: $a_i$ and $a_i^\star$. Therefore following the definitions of section \ref{subSec:extrinsic} on page \pageref{subSec:extrinsic}. More explicitly, the digital self connects to the intrinsic properties of data, while the physical self,$a_i$, connects to the extrinsic properties of data. 

For example, this separation clarifies some of the so-called "digital identities" questions. Real-world extrinsic properties of data define an "identity". These are related to the context in which the autonomous principal $a_i$ operates while only identifiers, an intrinsic data property, can relate to $a_i^\star$.

Suppose the information system design lets the autonomous principals have complete control over the confidentiality level of their outgoing messages. In that case, we can define a concept of trust in information systems that applies to the Principal-Agent question of this report
\par
{\sl Trust in information systems equals $a_i$ (i.e.Human) accountability + cryptographic assurance}
\par

To close this work, we restate our opening remarks of the introduction. This control over the confidentiality level is essential for information systems operating in challenging scenarios where the individual agent cannot rely on secure access to the information network (e.g., during natural disasters), is not protected by governance (e.g., in times of political instability), or cannot connect online (e.g., during economic distress). In such cases, safeguarding the digital self of the autonomous principal in the digital space is akin to protecting an individual in the humanitarian domain\footnote{as the Estonian government has done with the concept of digital embassy}. Consequently, the security of information systems also requires the existence of a digital safe harbour for autonomous principals in addition to the traditional cybersecurity elements..

%% file: CV4-Conclusions.tex
%
\section{Conclusions\\{\sl Facing realities}} 
\label{sec:conclusions}
{\bf Alternative governance models are needed.}\\
Information Theory became the stage upon which digital technologies developed when the theoretical basis of modern computing was laid out in the first half of the twentieth century through the ground breaking works of Turing \cite{Turing_1936}, J.von Neumann \cite{Gold_1980} and C.Shannon \cite{Shan_1948}.  As a result, almost 100 years after their publications, we live in a digitalised world globally. However, digital technologies also brought an unexpected side effect to their immense benefits to society: They are hard to govern. Why? The root cause is related to the fact that "Information" itself is hard to govern. "Information" spreads and does not have the unicity property of physical-world entities. Herefore, the primary governance concept of liability is hard to define. Adding the economic dimension to the governance problem led to technological architectures polarised toward centralised solutions. Zuboff's argumentation \cite{Zuboff_2019} becomes sadly a  logical outcome.   
\par
{\bf Governance leads to alternative digital transformation.}\\
This article introduced the distributed governance model (section 2) as an alternative mechanism for governing information exchange between economic agents. By addressing the governance problem through the perspective of the Principal-Agent question, the model introduces a definition of liability independent of whether the information exchange occurs in the physical or digital space. Furthermore, with digital technologies treated as the next stage of how humans exchange information, this approach subdues the impact of digital technologies on existing governance to secure societies. As such, the distributed governance model becomes an alternative digital transformation model. 
\par
{\bf A governance landscape based on autonomous principals, privacy spheres and ecosystems }\\
The model defines autonomous principals \ref{subSec:autonomousPrincipals} based on their intrinsic capacity for independent choice. Closely related to the concepts of free will or sovereign reason, this independence of choice is necessary to re-introduce transactional sovereignty in the digital space. Furthermore, as choice implies accountability of decisions, the distributed governance model cleanly assigns liability to an autonomous principal. Applied to the digital space, this demonstrates the necessity (and technical complexities) of a Digital-Self, a technological extension of the physical self.
The legal analytical concept of the privacy sphere is the basis for defining confidentiality spheres \ref{subSec:confidentialitySpheres} to provide a functional definition of "self" applicable to physical and digital spaces. As Figure \ref{subSec:extrinsic} on page \pageref{subSec:extrinsic} shows, the digital self must be intimately and unambiguously bound to the physical self. This connection becomes possible with the emergence of new decentralised software technologies in the domains of semantics and authentication. They are necessary for the autonomous principal to maintain control over his digital self without the interference of external parties. This perspective also requires a precise definition of Data vs Information (section \ref{subSec:extrinsic}) with a data-oriented network model instead of the traditional service-oriented (client-server) current models. This results in improved technologies for data lineage leading to mechanisms for identifying data provenance without creating centralised surveillance. 
\par
The governance results from the need for security demanded by a community of autonomous principals. Section \ref{subSec:ecosystems} defines the ecosystem for that purpose. An ecosystem is a population of autonomous principals gathered around a purpose represented by a legitimate authority. As such, autonomous principals, while keeping their integrity, implicitly consent to the rules the corresponding administration sets. This definition also implies that the same autonomous principal will be part of multiple ecosystems themselves interacting with each other. See section \ref{subSubSec:digitalPassports} for an example of the digital passport. Therefore, the ecosystem also becomes an autonomous principal when interacting with its peers. The definition of ecosystems can thus scale as each ecosystem becomes accountable for its decisions (e.g. bi-lateral agreement) that subsequently govern its population. 
\par
An immediate benefit of this approach is the confinement of algorithmic decision-making (A.I./M.L.) governance to the data network represented in figure \ref{subSec:extrinsic} (within the distributed governance model, algorithmic agents are considered non-autonomous even though they might be defined as autonomous in software engineering terminology). When implementing explicit instances of the distributed governance model with authentic data lineage, A.I. and other algorithmic decision-making processes are owned by individuals or organisations that become, as autonomous principals, accountable under domain-specific governance. 
\par
{\bf Implementation of distributed governance model}\\
The following steps are implementing the distributed governance model in real-world applications. Digital healthcare and digital transformation of the public sector are the primary application domains. Both deal with a potentially large population of distinct stakeholders and existing rules and regulations. 
The model is also being tested in the domains of supply chains with a particular focus on introducing IoT devices within a business process.
\par
The second part of this article will report on projects underway to apply the distributed governance model for developing health data spaces in cardiology and ontology in line with the European Health Data Space (EHDS) vision.
\par
{\bf Immediate action for long-term vision}\\
Our concern about the timely delivery of digital transformation in the humanitarian sector serving vulnerable populations was the starting point of our motivation (section \ref{subsec:humanitarian}). 
As re-stated at the start of this conclusion, the distributed governance model is designed to bridge governance between the human (i.e. physical) and digital space. As such, it can support our initial purpose by delivering digital governance burned into the active network protocols when the human governance in one ecosystem is incapacitated. Then, the four quadrants of Figure \ref{fig:caseForHumanitarian} on page \pageref{fig:caseForHumanitarian} will be secured, and our goal will be reached.
\par
The article's last section briefly introduces a trust infrastructure coined Dynamic Data Economy integrating the distributed governance model. On the technology side, one of its outcomes is to transfer from the application layer to the network layers' core security elements enabling what technologists call "fat protocols". For the end-user, standardised network protocols secure the information, thus relieving applications and platforms of many security burdens. 
This distributed and standardised view of the network protocols also leads to the global internet network governance concept as a common heritage of Humankind, as developed by Dr. Rolf Weber \cite{Weber:2022}.
\par
\subsection*{Acknowledgments} 
\label{subSec:acknowledgments}
This work was financed in part by the European Union Horizon 2020 research and innovation programme within the framework of the eSSIF-Lab Project funded under the grant agreement No:871932 and by the Human Colossus Foundation (HCF), a non-profit foundation for the development of the next-generation internet. The authors thank the HCF community of experts for their input. Noteworthy acknowledgement extends to Michal Pietrus for his profound insights and unwavering commitment to advancing the code underpinning the Dynamic Data Economy (DDE) framework.